\documentclass[doc,12pt]{apa6}

\usepackage[numbers,square]{natbib}

\newcommand{\bibTitle}[1]{``#1''}

\usepackage[colorlinks]{hyperref}
\usepackage{url}

\usepackage{caption}

\usepackage{multirow}

\usepackage{tabularx}
\newcolumntype{C}{>{\centering\arraybackslash}X} 

\usepackage{setspace}
\usepackage[utf8]{inputenc}

\usepackage{float}

\usepackage{amssymb}
\usepackage{amsfonts}
\usepackage{amsmath}

\usepackage{csquotes}
\usepackage{graphicx}
\usepackage{float}
\usepackage{comment}
\usepackage{booktabs}
\usepackage{subfig}

\makeatletter
\g@addto@macro\@floatboxreset\centering
\makeatother

\usepackage{afterpage}

\begin{document}

\author{Adnan Rebei}
\affiliation{University of Illinois at Urbana-Champaign, Champaign, USA}

\title{Entropic Decision Making}
\shorttitle{\footnotesize{Entropic decision making}}

\clearpage

\abstract{Using results from neurobiology on perceptual decision making and value-based decision making, the problem of decision making between lotteries is reformulated in an abstract space where uncertain prospects are mapped to corresponding active neuronal representations. This mapping allows us to maximize non-extensive entropy in the new space with some constraints instead of a utility function. To achieve good agreements with behavioral data, the constraints must include at least constraints on the weighted average of the stimulus and on its variance. Both constraints are supported by the adaptability of neuronal responses to an external stimulus. By analogy with thermodynamic and information engines, we discuss the dynamics of choice between two lotteries as they are being processed simultaneously in the brain by rate equations that describe the transfer of attention between lotteries and within the various prospects of each lottery. This model is able to give new insights on risk aversion and on behavioral anomalies not accounted for by Prospect Theory.

}

\thispagestyle{empty}


\maketitle

\setcounter{secnumdepth}{3}


\section{Introduction}

The underlying neuronal mechanisms of decision-making in the brain are sill largely unknown, especially in complex tasks where uncertainty in the choices is important in the decision process. However, at the behavioral level, risky decision-making models are numerous and vary in methods of treatment with little attention to the neuronal basis of the process. This work is an attempt to build a model of risky decision-making based on behavioral results that also accounts for some of the features we learned about the brain network during decision tasks.

The Allais paradox \citep{Allais1953} has been a driving force behind the development of utility-based theories for the last seventy years or so. The paradox was a concrete example that human behavior is incompatible with Expected Utility Theory (EUT) of  \citet{vonNeumann1943}. In spite of its simplicity, EUT has a concave utility function that successfully accounts for subjective values of goods and is consistent with the law of diminishing marginal utility as opposed to the naive linear expected value theory (EVT).
To study uncertainties in economic decisions, lotteries became a simple testing ground for theories of choice. In EUT, the prospect of a lottery is weighted by a linear probability which in a choice task between lotteries leads to common outcomes being ignored. This property is called independence of outcomes.  
The Allais paradox showed that the independence property of EUT is violated by most people.

Among the many extensions of EUT, Prospect Theory (PT), which is also an expectation-based model similar to EUT \citep{Kahneman1979,Tversky1992b}, is by far the most popular theory of decision making under risk. PT extended EUT by introducing subjective probabilities and accounting for loss aversion. The subjective probabilities are nonlinear functions of the probabilities of the prospects that give rise to inverse-S shapes to reflect non-linear behavior near small or large probabilities of outcomes.
 In PT, the decision maker (DM) is assumed to choose the lottery with the maximum utility  \citep{Fox2014}.

 Allais, in his 1953 paper, already suggested that linear probabilities in EUT may have to be modified along with including variance of the distribution of outcomes as an explicit parameter in the utility function. Rank-dependent utility theories, which cumulative Prospect Theory is an example, appear to show dependence on variance, but only indirectly through nonlinearities in the weights of prospects. While utility-based models of decision making are widely adopted, there are also other models that do not require a utility function. These models are 
  called heuristic models  and are only based on a set of simple rules \citep{Gigerenzer2011,Glockner2012}  that follow the "good enough" principle to make a decision \citep{Simon1955,Schwartz2002}. \citet{Simon1956} proposed in the 1950s that bounded rational decision makers do not commit to an
unlimited optimization by searching for the absolute best option. Rather, they follow a strategy
of satisficing, i.e. they settle for an option that is good enough in some sense. 
These heuristic models are not well developed analytically, but may prove to be better suited models in situations where the number of options is large or the probabilities of outcomes are unknown \citep{Gigerenzer2011}. These heuristic models were motivated by physiological considerations such as the limited cognitive capacity of a decision maker to make the best decision in each choice task. 
 Therefore, the principle of optimizing a utility function is not always the best way to make a decision in the face of uncertainty and risk \citep{Mousavi2014}.  In this paper, we adopt an approach that is both physically motivated and quantitative. It  takes into account the physicality of  the decision maker (DM) and that his or her  organism obeys the law of maximum entropy with some constraints \citep{Jaynes1957}. The optimization of entropy instead of utility has been previously suggested by  \citet{Candeal2001}. Their proposal was based on the similarities of the mathematical properties of utility and entropy if a 'preference' relation is defined on the state space according to possible transitions between the states or prospects.

 Allais considered lotteries with higher variance, but same mean, to be riskier. The direct association of variance with risk is a familiar concept in the modern portfolio theory in finance where  the efficient frontier of the portfolio is chosen by minimizing its variance  \citep{Markowitz1952}. But, rank-dependent utility-based theories do not take variance as an explicit parameter in the theory.
 One of the main reasons behind this neglect is that within PT, variance is minimized in risk averse situations such as the case when dealing with lotteries with positive outcomes; but variance is maximized in risk seeking situations such as the case of lotteries with negative outcomes. If a lottery with outcomes $x_1, x_2, ..., x_n$ and corresponding probabilities $p_1, p_2, ...,p_n$ are represented by $(x_1, p_1; \, x_2, p_2; ...., \, x_n,p_n)$, then  
   given two lotteries   $A^+ = (\$100, 1) $, and $B^+ =(\$500, 0.2;\,  \$ 0, 0.8)$, people tend to prefer lottery $A^+$ rather than lottery $B^+$ where lottery $A^+$ has zero variance. However, when it comes to choosing between lotteries $A^- = (-\$ 100, 1)$ and $B^-=(-\$ 500, 0.2; \,  \$0, 0.8)$, people tend to choose $B^-$ which has a larger variance. This asymmetry between positive and negative outcomes, called the reflection effect in PT,  was not exactly what Allais had anticipated in his original work. Because PT can explain this behavior of people avoiding sure loss, variance has not been explicitly incorporated in utility theories. 
   
   One of the main objectives of this paper is to include the variance of the distribution of outcomes explicitly in the analysis of risky decision making. Presently, computational models of decision making include some kind of variance in the analysis. This variance is  modeled as some additive noise that is added to each outcome. The noise is independently distributed across outcomes, and its strength is assigned ad-hoc. Theories that fall in this latter category include Stochastic Expected Utility, e.g, \citet{Blavatskyy2007}, and drift-diffusion-based models \citep{Ratcliff2015,Johnson2016,Busemeyer1992,Roe2001}.

The first attempt to use entropy as a measure of spread of outcomes in a lottery was by \citet{Meginniss1977}. He proposed the following extension to utility theory for a gamble $(x_1, p_1; x_2, p_2; ..., x_n, p_n)$ with utilities $u(x_i), i = 1, 2, ..., n$ and 
entropy $H(p_1,p_2, ..., p_n)$ \citep{Shannon1948} 
\begin{eqnarray}
U(x,p) & = & \sum_{i=1}^n p_i u(x_i) + \beta H(p_1, p_2, ..., p_n),
\label{meginnis1}
\end{eqnarray}
  where $\beta$ is a free parameter that can be either positive or negative.  \citet{Luce2008} and by \citet{Ng2008} developed this idea further axiomatically. Instead in this paper, our goal is to develop Meginnis' idea in  a more physical way.

In his thesis, Meginnis showed that his proposed utility function  satisfies transitivity, irrelevance of impossible outcomes, and  more importantly, that lotteries with equal probability weights are preferred to lotteries with different probability weights when prospects are the same. The extra entropy term allowed him to predict that risk-taking behavior should be observable in situations that involve large gains but small probabilities or a small loss with a large probability as is the case when buying a lottery ticket. Similarly, risk-taking behavior can be observed in events with small gains and large probabilities or  large losses with small probabilities as in the case of crossing the street to get a coffee. Another advantage of including an entropic term in the utility function is that he was able to show that buying insurance and gambling behavior are both still explainable with a monotonic concave wealth function as opposed to the idea that concavity is not a valid requirement in this situation \citep{Friedman1948}. 

 To explain the Allais paradox, Meginnis needed a negative scaling parameter $\beta$      in Eq. \ref{meginnis1}. This signals that in the Allais lottery, the DM is entropy averse. A negative $\beta$ was also needed to get a similar efficient frontier as in  the modern portfolio theory of \citet{Markowitz1952}. In this paper,  the parameter $\beta$ is fixed by the distribution of prospects independent of other lotteries. However, this is achieved only if variance of the distribution of outcomes is also included as an additional term in the utility. In the theory presented here, the explicit inclusion of variance  amounts to controlling the convexity of the decision maker's utility function. Experimental behavioral economics and neuroscience data support the idea that a DM is sensitive to variations in variance of choices.
   For example, in the ultimatum game, it was demonstrated by \citet{Vavra2018} that both the mean and variance of offers from the proposer have direct and different bearings on the response of the receiver. A smaller variance makes the receiver less accepting of "unfair"  proposals, while a larger variance showed a more relaxed pattern of accepting or rejecting offers which can be interpreted as a noisier response than in the low variance case.  Sensitivity to variance has been also observed by neuroimaging techniques. \citet{Kobayashi2010} measured different responses in orbitofrontal neurons in animals to two distributions of juice volumes with the same mean but different variance.  When processing various images, retinal cells were also shown to adapt to variance in a stimulus to efficiently process a wide range of light intensities \citep{Smirnakis1997}. Variance can also have a discounting effect on rewards. In a rapid serial visual presentation paradigm, \citet{Apps2015} showed that participants were  affected by both the amount of cognitive effort exerted in a task, which is measured by the number of shifts of spatial attention, and to the variance in the possible amount of cognitive effort that need to be exerted. Both factors were found to be independent. This shows that variance can act as a cost function in a benefit-cost type of analysis and can affect decision making independently of other factors.

 In spite of the successes of PT, decision making models in risky situations still remain a work in progress. An extensive replication study of the behavioral phenomena that supported PT was recently conducted by \citet{Millroth2019} with a broad spectrum from the population as opposed to using only students. They found that the replication of the new predicted behaviors by PT was poor. From the nine paradoxes that PT explained, more than 50 \% of the participants exhibited at most  three of the nine paradoxes studied by \citet{Kahneman1979}.  Similarly, the functional form of PT is not unique. \citet{Stott2006}, who extensively studied the possible different forms of PT in the literature, concluded that the weighting function of subjective probabilities is best described by Prelec's exponential-log form \citep{Prelec1998} rather than the initial algebraic form suggested by Kahneman and Tversky. In general, PT requires three parameters to capture the distortion of probabilities, the curvature of the value function, and loss aversion. In addition to these parameters, there are some editing rules like fusing or coalescing branches with similar outcomes that need to be applied to the gamble before fitting it to the PT model. PT itself has two versions, one that violates stochastic dominance and another that does not.  Stochastic dominance is a fairly standard property of preference relations obeyed a `rational' DM which basically expresses that more is better, but this property requires the cumulative probability distribution function of outcomes to express it mathematically, a non-trivial matter to  carry out consciously by a DM. In fact, it is now believed that stochastic dominance can be violated by a DM and that this violation is genuine and not due to framing effects \citep{Loomes1992,Birnbaum2004b,Dertwinkel2015}. Attempts to provide a decision theory that accounts for violations of stochastic dominance have already appeared in the literature \citep{Birnbaum2004b,Dertwinkel2015,JohnsonLaird2006}. In this paper, we will also provide a theory that allows for stochastic dominance violations, but the motivations for this violation are different. They are mainly an interplay between averseness to entropy and risk.

 In addition to violations of stochastic dominance, PT was also found to violate preference behavior in situations where lottery branches are coalesced or split in a particular way irrespective of framing effects \citep{Birnbaum1996,Birnbaum1997,Birnbaum1998,Birnbaum2004a,Birnbaum2004b,Birnbaum2008,Birnbaum2018}.
   To address these new paradoxes, Birnbaum proposed a new descriptive theory, the transfer of attention exchange model (TAX), that accounts for old and new violations. While TAX did not model attention in any physical way, we will provide in this paper an explicit model of attention guided by working memory (WM) and based on a Markovian process. The memory part of the model plays the role of bookkeeping for  the mental state similar to the role played by a Maxwell demon \citep{Maxwell1891}. In 1929, \citet{Szilard1929} provided the first example of a Maxwell demon that integrated physical entropy with information entropy by making the memory of the demon part of the physical system. In short, the Maxwell demon plays the role of a 'traffic controller' that uses gained information to extract work from the system \citep{Parrondo2015}. By analogy, the demon or WM  interacts with  lotteries to gain information that can be used to maximize anticipated utility.  Thus our intent is to develop ideas that can be extended to deal with real ensemble of neurons starting from behavioral data. This is the reason behind insisting on a formulation that respects physical laws. However, this requires a new reinterpretation of the gamble in terms of biological inputs which necessarily includes an energy component in addition to an entropic contribution.

 Since the human brain is a physical system, it must also satisfy established physical laws, such as that of maximizing entropy,  while processing information to produce a decision.  It is important to keep in mind that a decision is a process that is inherently irreversible and involves global changes in the mental state. According to  neuroimaging studies using fMRI, the brain is modular which implies that decisions are most probably made within a specific set of neurons and then the outcome of the  decision is propagated to other areas of the brain responsible for action.  It has been so far difficult to pinpoint a specific set of neurons that are always responsible for decision making, but what is known so far  is that  areas in the cortex such as vmPFC, posterior cingulate cortex (PCC),  striatum are involved in economic decision making \citep{Clithero2014,DeMartino2009}. Other areas that partake in the decision may  include LIP \citep{Roitman2002}, dorsolateral PFC \citep{Sanfey2003}, and basal ganglia \citep{Bogacz2007}. The take-away point from these studies is that the instant the decision is made, it is locally executed in the brain. Hence during that window of time, the subset of neurons that are directly involved in the transition from a multi-option state to a single state can be considered to be strongly interacting in the presence of a background of other neurons from other areas of the brain. For example, in economic decision problems, the work of \citet{Padoa2017} points to the OFC region of the cortex as the most probable region where the decision is being executed. The local character of computations in the brain is supported by the organization of the neural network itself. Compared to the available local connections, there are only sparse long-range connectivity in the cortex. It is believed that this structure may be advantageous to have an efficient computation system where intensive calculations, like a decision, are carried locally and only the result is transmitted to other areas of the brain \citep{Laughlin2003}. This will be an important assumption in our model that we will use to motivate our entropic formulation of the decision process.
 \begin{figure}[!htb]
 	\includegraphics[width=8in,height=6in]{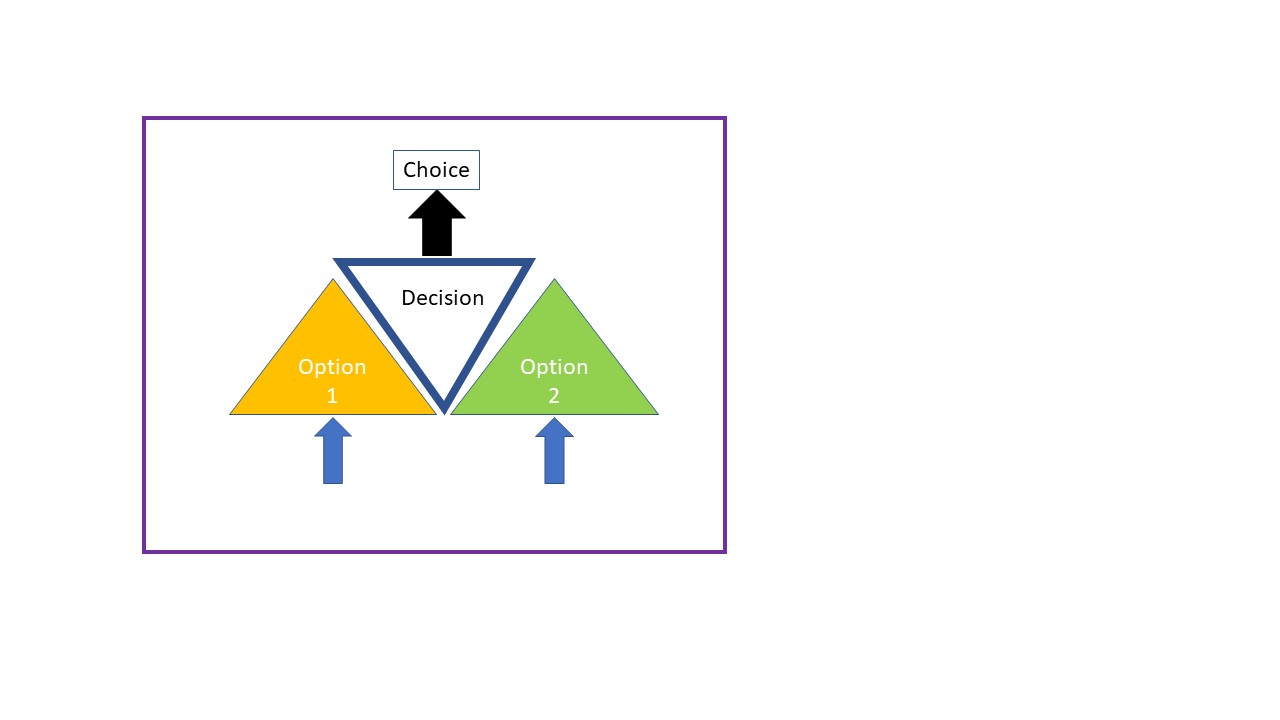}
 	\caption{Neural representation of a two-choice task. Each choice is initially represented by a group of neurons independently of the others. During the decision phase, both sets of neurons become coupled. Both sets of neurons directly involved in the decision state reside in the cortex. }
 	\label{neuralrep}
 \end{figure}
 
   Pictorially, the neurons that are encoding two prospects are represented by the neural network blocks in Fig. \ref{neuralrep} for a two-choice task. Right before the decision phase, each group of neurons representing a prospect can be considered in a metastable state since the neurons in that set are mostly strongly interacting with each other and with relatively weaker interaction with the rest of the neurons in the brain. Given that each neuron interacts approximately with about $10, 000$ other neurons, the neurons representing prospects 1 and 2 can be considered in a quasi-equilibrium state independent of those representing the other prospect(s).  Since we are mainly interested in value-based decision making behavior, we will assume that value is expressed in terms of the intensity of firings or interactions in the neurons that are encoding a prospect. This is a generally valid assumption   used in popular models of neuronal dynamics. 
In Wang's model of motion perception of parietal neurons, higher firing rate corresponds to higher value \citep{Wang2002,Wong2006,Wang2008}. In the Good-based model of \citep{Padoa2011}, higher firing rates are also correlated with higher values. The neuronal representation of the prospects that we adopt in this paper were motivated by measurements of Padoa-Schioppa's group \citep{Padoa2013,Padoa2014,Padoa2006}.

Finally, we need to point out one important discrepancy that exists between microscopic (neuronal-based) and macroscopic (behavioral-based) models in perceptual decision making. According to the microscopic treatments, winner-take-all network models best describe the decision process in which the two choices are represented in
the firing rates of separate populations of neurons. The populations compete through
interneuron-mediated inhibition leading to a winner-take-all behavior \citep{Wang2002}. On the other hand, most currently popular macroscopic models rely on an accumulating evidence procedure to reach a threshold  upon which a decision is made
 \citep{Roe2001,Usher2001,Trueblood2014}. There is still no consensus if the mechanism for perceptual decision is the same for value-based decisions \citep{Gold2007}. We will  assume  that value-based decisions and perceptual decisions share similar dynamic features in brain regions where the neuron populations of discarded prospects occur.

The rest of the paper is organized as follows. We first introduce a simple experiment of two gambles that have close to extreme probabilities and equal expectation values. Our measurements give results that are in direct opposition to Prospect Theory and the TAX model. By treating the decision problem as a signal processing problem, we show that our simple model gives the measured results. Our model also does not distort probabilities or utilities but 
 assumes fluctuations in the number of spikes in the population of neurons encoding the prospects, i.e. noisy probabilities, and maximizes the Shannon entropy of the gamble. This method has no free parameters and  has parallels with signal detection theory \citep{Zenon2019}. Next, we extend Meginnis' method by using non-extensive entropy and adding variance explicitly to the 'utility' function. We apply this new model  to the equal expected value lotteries example.  This model has only a single parameter that is interpreted as due to the interaction of two choices in the brain of a  DM. The model is able to explain the new anomalies discovered by  \citet{Birnbaum2004b}. To account for some of the dynamical features of a decision task in the brain, a theory of Attention is proposed. This extended model is constructed by analogy with the Maxwell demon in information engines. It  models bias and (implicit) attention switching rates between  all prospects of both lotteries in a two-choice task. The model allows us to estimate the percentage of population choosing one lottery and not the other. All three approaches that we discuss are complementary and should be part of a comprehensive theory of decision making. 
 In the final section, we summarize our results and briefly discuss extensions of the ideas presented in this work.

 \section{Decision Making as a Signal Processing Task: The case of equal expected values} 

One important finding in  neuroeconomics is that subjective values are represented explicitly at the neuronal level \citep{Glimcher2014}. In a decision making task, different cells  encode different variables. \citet{Padoa2011} reported that the cells that encode values and choices are found to be in the orbitofrontal cortex (OFC) region of the brain. However, background noise in the OFC is significant and may lead to fluctuations in the encoded variables such as the probabilities of the lotteries. In this section, we will consider only this effect in a very simplified model of neuronal excitations.

The case of equal expected values and equal entropies provide a stringent test of the ideas proposed by Meginnis \citep{Meginniss1977}. To be concrete let's consider the following case inspired by studies from the work of \citet{Lichtenstein1971} on bets with high probability values and low utility values of outcomes and vice-versa. As an example, let's consider the following gambles $A = (\$ 10, 0.90; \$ 0, 0.10)$ and $B = (\$ 90, 0.10; \$ 0 , 0.90)$.  Here, we have $EV(A) = EV(B) = \$ 9$, where EV stands for expected value, and $H(A) = H(B) = 0.325$. Using $u(x_i) = x_i$ in Eq. \ref{meginnis1}, we have the utility of both gambles $U = 9 + 0.325 \beta$ . Therefore, in this case there is no $\beta$ that can give a preference order between both gambles. However, the majority of people prefer gamble $A$ to gamble $B$.  Cumulative Prospect Theory \citep{Tversky1992b} is also of no help in this pair of gambles unless the probability weight function is  allowed to be 'S' shaped, i.e., $\gamma > 1$,  which is not supported by measurement.  Using their value $\gamma = 0.61$ for their original  probability weight functions \citep{Tversky1992b}, we find that $U(A) = \$ 7.12$ and $ U(B) = \$ 16.77$. Therefore, according to PT, $B$ is preferred to $A$.   For the TAX model of \citet{Birnbaum2004b}, we find that $U(A) = \$ 5.49$ and $U(B) = \$ 10.6$. Therefore, similar to PT, the TAX model predicts that people should choose $B$ over $A$ \footnote{These numbers for the TAX model were calculated using the online calculator available at Birnbaum's site:  http://psych.fullerton.edu/mbirnbaum/calculators/taxcalculator.htm}.

To get around these difficulties, we instead adopt a more microscopic (biophysically-inspired) picture of the process of decision making. As we noted in the previous section, probabilities are positively correlated with firing rates between neurons. In a simple perceptual decision task that consists of identifying the direction of motion of a dot on a screen, neurons in the middle temporal (MT) of the brain, which are sensitive only to motion directions, start firing with higher frequency consistent with the behavioral response. Moreover, fluctuations in the firings are also reflected in fluctuations in the decisions \citep{Britten1996,Nienborg2009}. 
Adopting this picture, we can easily see that because of the noisy environment in the brain, there will be a distribution of firing rates representing the probability distribution of outcomes themselves. In other words, the probabilities of outcomes represent only an average of a distribution with some variance. These fluctuations in the probabilities may become important in a pair of gambles like the one we are considering. Of course, we may expect also fluctuations in the representations of the values of the outcomes too. But for the moment, we will confine ourselves to try to account for the fluctuations around the probability values. In principle, a microscopic treatment should be able to provide a probability distribution of outcomes and  any fuzziness around those values.

In this paper, we are going to focus our attention on the state of neurons after the encoding phase of the stimulus and right before the actual physical decision is carried out on the neuronal level, i.e., when the various neuronal assemblies representing the various options are transformed to a single assembly representing the chosen outcome. 
To keep a faithful representation of the decision task, the neuronal cells must maintain a stable firing rate for each outcome. In this example, the gamble has two outcomes which will be represented by two neuronal populations, eg., in the OFC region \citep{Padoa2006}. It is believed that it is this phase of the decision process that is the most energy-intensive and that the magnitude of neuroimaging signals reflects this processing, while neurons in the post-decision phase appear to be less active \citep{Harris2010}.  This is also supported by observations made on  neuronal activities in monkeys after the external cues have disappeared from the view \citep{Durstewitz2000}. Given the large number of neurons in the frontal area of the cortex, about 10 billion \citep{Pelvig2008}, and a large number of synapses per neuron, this state, just before the decision, can last several hundred milliseconds and stay coherent \citep{Garrido2007} during that time. This implies some sort of a metastable state is established before a signal 
 from neurons carrying the output is registered.  While we will be interested in this particular state, the details of how the output comes about is outside the scope of this paper. However, the scenario we adopt here is more in line with a winner-take-all paradigm rather than an accumulation one,  and the spikes are assumed to maximally distribute themselves among the different option states of the gamble. Presently, diffusion models, which are based on evidence accumulation, are more popular because of  their flexibility to model  many perceptual decision behaviors \citep{Ratcliff2016,Roe2001,Usher2004}, but there are other models that do not rely on this scheme that evidence is being accumulated until a threshold is reached \citep{Bogacz2007,Gurney2001,Shadlen2013,Caballero2015,Caballero2018}.  It is also important to remember that the diffusion picture is not supported by what happens at the neuronal dynamic level \citep{Roxin2008} which at best, under crude assumptions, can be reduced to a nonlinear diffusion equation in the case of a two-choice decision. Therefore, our broad assumptions are justified since they don't put severe constraints on the actual dynamics of neuronal spiking. 
 
 Next we discuss one last ingredient in the proposed model, i.e., energy. Many studies in the literature consider decision making as information processing in the brain. Therefore, the goal in those studies is to find an efficient mechanism to carry out this process. However, information processing cannot be disentangled from the process of energy consumption. The brain consumes almost 20\% of metabolic energy per weight continuously \citep{Hofman1983}. Moreover, information processing is costly. It takes about      
 $10^4$ ATP molecules to transmit a bit at a chemical synapse and $10^6 - 10^7$ ATP molecules for
 graded signals in an interneuron, photoreceptor, or for spike coding \citep{Laughlin1998}. In a study on photoreceptors in flies, it was observed that higher light intensities require much more energy to process. Therefore, energy 
 considerations can lead to stricter constraints than those due to 
  information processing \citep{Niven2007}.  Other studies also showed that taking into account energy considerations in addition to neural coding efficiency changes the understanding of what is being optimized. Consideration based solely on information theory tend to have a goal of minimizing entropy \citep{Kelly1956,Barlow1961}. However, when energy considerations are included, efficient information processing, i.e., firing rates,  can sometimes favor increases in entropy \citep{Levy1996}. According to \citet{Padoa2011}, each neuronal response encoded only one variable, and the encoding was linear. Therefore, higher outcome values translate into higher firing rates which implies higher energy consumption. In other words, the higher the utility function, the higher the energy needed to represent that utility in the brain. This is how energy is introduced in our model which may translate into maximizing motivational states or anticipating states rather than gains. Therefore, an outcome in a gamble will be considered as an abstract energy state among other available energy states representing other outcomes. Neurons themselves are not important in this picture; only the different populations of spikes and their number is what interest us here. Throughout the lifetime of the metastable state before the decision, spikes are created and destroyed through feedback loops, but the average number of spikes at any instant\footnote{The term `instant' here means a time interval that is very short with respect to the time scale of the decision which is of the order of 100 $msec$.} in time is assumed to be on average constant in each population. If we treat each gamble separately, it is abstractly represented by a system of `energy levels' that represent value `populated' by corresponding spikes  with their corresponding relative number representing probabilities. Taking this point of view, our decision problem becomes an inverse problem of a familiar problem in statistical physics that of determining the distribution of particles among $n$ energy levels in contact with a heat bath at given temperature \citep{Huang1987}. In our case, we have a distribution of spikes over $n$ outcomes, and we want to determine external parameters due to other neurons in the background that help in maintaining the metastable state of the neurons directly involved in the decision phase. This model is similar in spirit to the Good-based model of \citet{Padoa2011} and \citet{Padoa2017} in that the states of the decision space are labeled by the outcome values. This model, as it stands, is incapable of leading us dynamically to the post decision state, since the interneurons in the OFC that encode the chosen value are not part of the model \citep{Rustichini2017}. But in the presence of another gamble, we will assume that the gamble that is less noisy will be the one that will be transformed to a chosen state given that the average strength of both signals (i.e., gambles) are the same. This is what we will show next.

\begin{figure}
 	\includegraphics[width=8in,height=6in]{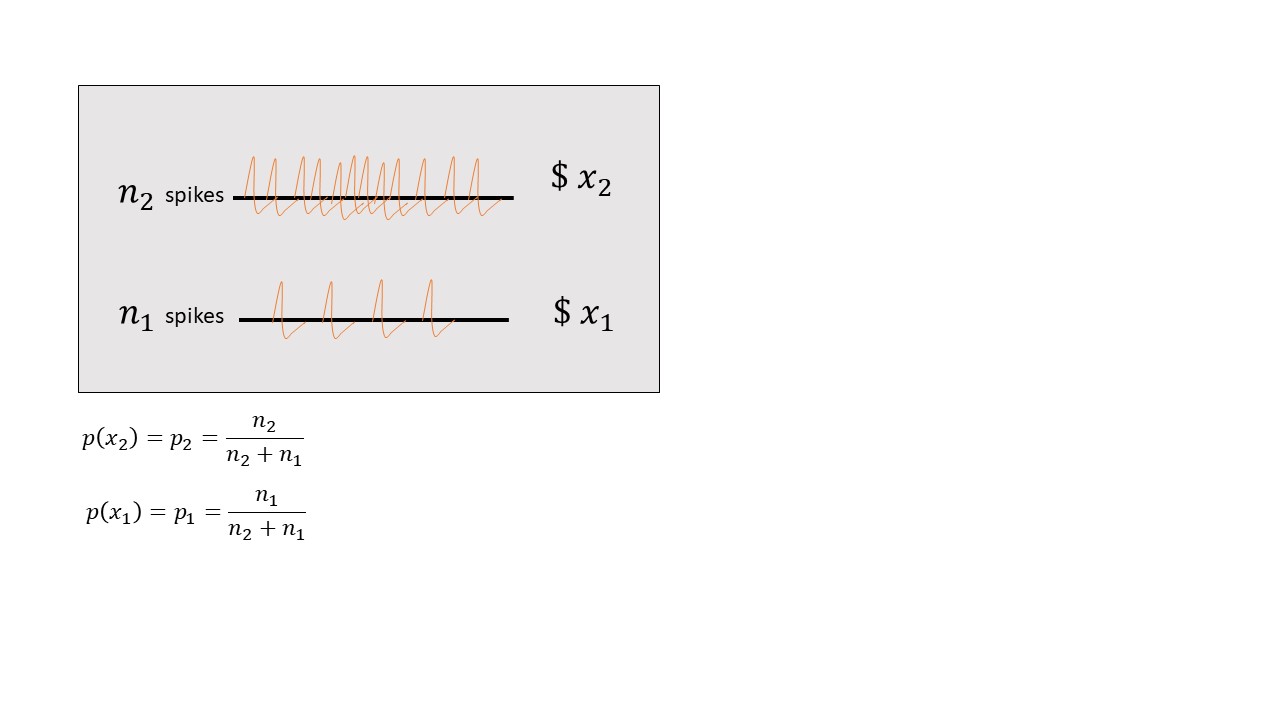}
 	\caption{A schematic representation of the metastable state of the neurons representing the stimulus in the decision phase and right before the 'birth' of the chosen state. The spikes 1 and 2 are only differentiated by their frequency and the population of neurons that form the background for these spikes. These neurons themselves are in contact with another large number of neurons that may be part of the cortex or subcortex, such as the amygdala, that are considered part of the ambient medium or bath.}
 	\label{slovic2}
 \end{figure}

 In Fig. \ref{slovic2}, a pictorial representation of the abstract space of stimulus representation by neurons in the cortex is shown. Given that all the spikes are similar, the overall state of the gamble is therefore symmetric with respect to an exchange of two spikes irrespective if they encode outcome 1 or outcome 2.  We will assume in the following that outcome $x_2 > x_1$. 
 Therefore,  the metastable state is characterized by two parameters,  the average energy, $ \bar{u} = p_1 x_1 + p_2 x_2$ (i.e., utility) and the average number of total spikes $n_s = n_1 + n_2$. The averages are defined with respect to the exponential (Boltzmann) distributions that maximize the Shannon entropy and are given by \citet{Huang1987}: 
 \begin{eqnarray}
 p(x_i) & = & \frac{1/n_s}{\lambda e^{x_j/\beta }-1}
 \end{eqnarray}
 Since the average $p_i = n_i/n_s$, then any noisy encoding of the stimulus will lead to noisy encoding of the anticipated utility, $\overline{\Delta u^2} = \overline{u^2} - {\bar{u}}^2$. The signal to noise ratio, SNR, for this gamble is therefore equal to, $SNR = \bar{u}/\left( \overline{\Delta u^2} \right)^{1/2}$, or
\begin{eqnarray}
SNR & = & \frac{p_1 x_1 + p_2 x_2}{|\beta| \left[ \left(p_1^2+p_2^2 \right) \ln \lambda ^2 + 2 \left(\frac{1}{n_s}\right)^2 \right]^{1/2}} 
\end{eqnarray}
where
\begin{eqnarray}
 \beta & = & \frac{n_s p_1 p_2}{p_1 - p_2} \left( x_2 - x_1 \right), \\
\ln \lambda ^2 &= & \left( \frac{x_1}{\beta} - \frac{1}{n_1} \right)^2.
\end{eqnarray}
It should be noted that the SNR is zero when $p_1 = p_2 = 0.5$, which is what we expect if only spikes determine the outcome as assumed here.
 
For our particular gambles, we have  for gamble A, $x_1 = 0, x_2 = \$ 10, p_1 = 0.10$ and $p_2 = 0.90$, which amounts to $SNR(A) = 0.873$. For gamble B, we have $x_1 = 0, x_2 = \$ 90, p_1 = 0.90$ $p_2 = 0.10$. In this case, $SNR(B) = 0.512$, which is less than   the signal-to-noise ratio of gamble (stimulus) A. Hence, treating choice between gambles A and B as an information processing problem, it is more efficient to process A, the gamble with lower variance, than gamble B. This result agrees with what most people choose, and it does not agree with the predictions of PT and the TAX models. Therefore, for equal utilities, it is shown here that variance in outcomes becomes the decisive factor in the decision making process. Interestingly, the same conclusion has been reached in a perception experiment using the more elaborate theory of  Friston's minimization of surprise \citep{Schwartenbeck2015}.
Hence, we turn next to formulating our choice problem in terms of maximization of entropy with  constraints that include variance of outcomes explicitly.

\section{Decision Making from Non-extensive Entropy Maximization}

The discussion in the previous section was based on adopting a simple representation of the underlying environment of the spikes that encoded the stimulus. To address more complicated situations, we will instead apply Jaynes' principle of maximum entropy \citep{Jaynes1957} to the macroscopic states but include risk or variance of outcomes explicitly as suggested by \citet{Allais1953}. Moreover, we will use a generalized form of entropy, non-extensive entropy, to describe the mental states of a DM. 

\citet{Meginniss1977}, in the Appendix of his thesis, pointed to the possibility of using non-extensive entropy instead of Shannon entropy. Non-extensive entropy was first suggested by \citet{Havrda1967} based solely on mathematical properties of entropy when the additive property is no longer valid. The first serious applications of non-extensive entropy appeared in physics \citep{Tsallis1988} and later by \citet{Ng2008} to decision making. Here, we will follow our treatment in the previous section, but start directly from maximizing the non-extensive entropy under constraints of both average outcome and variance of the gamble. These constraints can be understood in terms of the adaptability of neurons to stimuli intensity and number of outcomes. 
The idea of maximizing entropy, adopted here for value-based decisions under uncertainty,  is similar to  the one used by \citet{Schwartenbeck2015} in the analysis of a perceptual decision task. The maximization of entropy in their case is reinterpreted in terms 
 surprise minimization, i.e., minimize differences between likely outcomes and desired outcomes \citep{Friston2009}. Formally, this can be written as a minimization of the Kullback-Leibler divergence between the DM prior preferences and posterior beliefs about the likely outcomes given current observations. From a behavioral point of view, the DM is keeping all options open in order to be least surprised with the outcome. 
Similarly here, we can also argue that maximization of entropy is a reflection of the DM to keep 
 all options available for an accurate evaluation of the gamble, but instead we will adhere mostly to our previous energy-based (i.e., value) picture.

 The use of non-extensive entropy, instead of Shannon entropy, allows us to treat a variety of paradoxes \citep{Birnbaum2008} that Prospect Theory cannot account for them. From a physical point of view, using non-extensive entropy is more in tune with the microscopic dynamics of neurons that gives rise to power-law distributions than the exponential Boltzmann distribution \citep{Alemany1994,Miller2006}. Therefore, the q-parameter of the non-extensive entropy
 \begin{eqnarray}
 H_q & = & \sum_i \frac{p_i^q - p_i}{1-q},
 \label{entropyx}
 \end{eqnarray}  
can be linked to intrinsic features of internal dynamics of neuronal networks, such as fluctuations driven by the stimulus \citep{Biro2005} or interactions between lotteries in a choice task.

\subsection{The Model}

Instead of a full-blown neurocomputational model, we aim to propose a model that conceptually bridges both behavioral and neuronal approaches. This has three aims. First, it adopts the behavioral responses as the `stuff' to be explained. Second, to build a model that is biophysically plausible and able to describe the macroscopic properties of spiking neurons.  Third, the model must  be consistent with physical laws of non-equilibrium thermodynamics. 

The structure of the neuronal network reflects functional specialization (or modularity) and functional integration of the brain. Modularity allows us to treat a specific phase of a cognitive task locally while averaging the effect of the rest of the brain on this local population of neurons. In the visual system, the most widely studied system, neurons in the visual area V5 are very homogeneous and are only specialized in detecting motion. In the decision making problem, neurons in the OFC are believed to be responsible for input, output value and chosen good processing \citep{Padoa2006}. Therefore, neurons in this area will be our focus, while the rest of the neurons in the  brain will be represented by parameters $\beta$ and $\lambda$.

For any gamble $G = (x_1,p_1; x_2, p_2; ..., x_n, p_n) $, a corresponding generalized
Meginnis' utility is obtained by maximizing the non-extensive entropy using q-weighted averages, Eq. \ref{entropyx},  
subject to the conditions 
\begin{eqnarray}
\label{constraint1q}
\sum_i \; p(x_i) & = & 1,\\
\label{constraint2q}
\frac{1}{\sum_i p(x_i)^q} \sum_i \; p(x_i)^q \; \left( x_i - x_M \right) & = & 0, \\
\frac{1}{\sum_i p(x_i)^q}\sum \; p(x_i)^q \;(x_i- x_M)^2  & = & \sigma^2,
\label{constraint3q}
\end{eqnarray}
where $x_M$ and $\sigma$ are fixed real quantities that may be used to enforce value normalization in the response of neurons to a stimulus \citep{Rangel2012}. To keep our treatment simple as in the previous section, we assume we are in a linear regime where value encoding is linear, i.e., $u(x_i) = x_i$.  Similarly, we will assume that there is a linear relationship between energies needed (e.g., in the form of firing rates) to keep the gamble representation active in the brain and corresponding outcomes.   This representation is supported by neurobiological findings. For example, \citet{Kennerley2009} found a sizable population of OFC neurons are modulated by three variables: the juice quantity, the action cost, and the probability of receiving the juice at the end of the trial. The firing rate increased with quantity and probability, as in the \citet{Padoa2017} study. Therefore, (positive) value is understood in terms of energy expanded to represent a positive outcome, or the motivation toward acquiring that outcome. Hence, an approximate variational principle can be written for this `coarse-grained' energy representation of the gamble as follows \citep{Martinez2000}
\begin{eqnarray}
\delta \left[ H_q(p) - \alpha \; \left(\sum_i \; p(x_i) - 1\right)  \right. &- \beta & \left. \left(\sum_i \; x_i \; p(x_i)^q - x_M \right)/\sum_i p(x_i)^q \right.  \\
& - \lambda & \left. \left(\sum_i (x_i-x_M)^2 \; p(x_i)^q - \sigma^2 \right)/\sum_i p(x_i)^q \right]  \; =  \;0. \nonumber
\label{optimizationq}
\end{eqnarray}
The parameters $\alpha$, $\beta$, and $\lambda$ are the associated Lagrange multipliers of the constraints, Eq. (\ref{constraint1q}-\ref{constraint3q}). From a behavioral perspective, the parameter $\beta$ determines the average utility while $\lambda$ is correlated with the variance in outputs or the amount of risk involved in the gamble. The parameter $q$ reflects the degree of deformation of subjective probabilities compared to those in the stimulus, or a parameter reflecting  long-range interactions between the neuron populations associated with  the various outcomes. The limit $q =1$ is the case treated in the previous section and represents independent neuron populations. Therefore, from a dynamical point of view, on the neuronal level, the q-deformations are a signal that there are long-range correlated fluctuations in the firing rates of the various populations. This is what we expect if coherence is maintained among neurons that encode the same outcome in a gamble and between choices \citep{Rustichini2015}. It should be apparent from our formulation that we are not addressing how the stimulus is being processed from the environment. This is a complex step that will involve finding the true distribution of states (outcomes) of the neuronal network driven by the external stimulus $s$, $p(x|s)$. In this case, the variational principle will involve a negative KL-divergence, $\mathcal{D}_q(p|p(x|s))$, between the distribution $p(x)$ and the actual one. Instead, we are assuming that all the microscopic states of the neuronal network are equally likely to be accessible. This is the microscopic origin of keeping all options open by the DM. 

If we rewrite the constraints in the simpler form
\begin{eqnarray}
\sum_m p_m & = & 1, \\
\frac{ \sum_m p_m^q u_m}{\sum_m p_m^q} & = & \mathcal{U}, \\
\frac{ \sum_m p_m^q u_m^2}{\sum_m p_m^q} & = & \mathcal{V}.
\end{eqnarray}
The solution of this optimization, Eq. (\ref{optimizationq}), gives \citep{Martinez2000}
\begin{eqnarray}
p_m & = & \frac{1}{Z_q} \left[  1 -(1-q) \beta^* \left( u_m - \mathcal{U} \right) - (1-q) \lambda^* \left( u_m^2 - \mathcal{V} \right) \right]^{\frac{1}{1-q}}
\label{psolution}
\end{eqnarray}
where the normalized parameters 
\begin{eqnarray}
\beta^* & = & \frac{\beta}{\sum_m p_m^q}, \\
\lambda^* & = & \frac{\lambda}{\sum_m p_m^q},
\end{eqnarray}
and
\begin{eqnarray}
Z_q & = & \sum_m \left[  1 -(1-q) \beta^* \left( u_m - \mathcal{U} \right) - (1-q)\lambda^* \left( u_m^2 - \mathcal{V} \right) \right]^{\frac{1}{1-q}}.
\end{eqnarray}
Using the solution of the variational principle, Eq. \ref{psolution}, in the entropy, Eq. \ref{entropyx}, we find that

\begin{eqnarray}
\beta \frac{\partial \mathcal{U}}{\partial \beta}   & = &   \frac{\partial}{\partial \beta} H_q \left[ p  \right], 
\end{eqnarray}
and
\begin{eqnarray}
\lambda \frac{\partial \mathcal{V}}{\partial \lambda}   & = &   \frac{\partial}{\partial \lambda} H_q \left[ p  \right].
\end{eqnarray}
Both of these equalities lead to the following relations
\begin{eqnarray}
\frac{\partial H_q \left[ p   \right]}{\partial \mathcal{U}}  & = &   \beta ,
\label{beta_deriv} 
\end{eqnarray}
and
\begin{eqnarray}
\frac{\partial H_q \left[ p   \right]}{\partial \mathcal{V}}  & = &  \lambda.
\label{lambda_deriv}
\end{eqnarray}
A new functional, $\mathcal{C}$, that is optimal with respect to utility $\mathcal{U}$ and risk $\mathcal{V}$ is defined as follows
\begin{eqnarray}
\mathcal{C} \left[\mathcal{U}, \mathcal{V} \right]  & = & \mathcal{U} + \frac{\lambda}{\beta} \mathcal{V} - \frac{1}{\beta} H_q \left[p  \right].
\end{eqnarray}
Using Eqs. (\ref{beta_deriv},\ref{lambda_deriv}), we can easily see that 
\begin{eqnarray}
\frac{\delta\mathcal{C}}{\delta \mathcal{U}} & = & 0, \\
\frac{\delta\mathcal{C}}{\delta \mathcal{V}} & = & 0.
\end{eqnarray}
Therefore, this is the new function that controls decision making processes in the mind of the DM, where the parameters $\beta$ and $\lambda$ control the average value and the spread of the options, respectively. These parameters are fixed by the distribution of outcomes in the stimulus, and possibly the state of mind of the DM.
The decision function $\mathcal{C}$ is explicitly given by
\begin{eqnarray}
\mathcal{C} & = & \sum_m \frac{p_m^q}{\sum_l p_l^q} \left(\frac{\lambda}{\beta}u_m^2 + u_m  \right) -\frac{1}{\beta} H_q \left[ p \right].
\end{eqnarray}
The function $\mathcal{C}$ is not necessarily convex, and that depends on the sign of both parameters $\beta$ and $\lambda$. Hence for negative ratio $\lambda/\beta$, the functional may have only maximum values. This is the case we are interested in if $\mathcal{C}$ is to play the role of a generalized utility function. This $\mathcal{C}$ function will be called an entropic utility. It must be understood that this is not an actual utility, but more like a motivational effort function. To understand this, we recall that we divided the brain into two parts; one small part that includes the actual neuron populations just before a decision is made, and a larger part, the rest of the brain, which constitutes in our picture the environment. In essence, we are considering only the delayed stimulus response only in the period just prior to the choice state. If we consider the utility part of the $\mathcal{C}$ as the 'energy' needed to represent the gamble states, then negative $\mathcal{C}$ may be considered from the point of view of the environment the difference between the  'information' cost  needed to maintain  the small population in their states by the rest of the neurons in the brain. Therefore, it makes sense to minimize this difference and choose the gamble that requires the least-effort to maintain, which is tantamount to a cost-benefit analysis. It has been suggested that a function such as the negative entropic utility is combined in a single signal in the brain that compares anticipated benefit and relative costs \citep{Basten2010,Pezzulo2013}. In an attempt to measure mental effort in an arithmetic addition task, it was observed that effort and anticipated reward share the same neuro-cortical areas \citep{Vassena2014}. The finding that high effort and high reward both elicit strong firing rates in the ACC and striatum was not expected but is 
in-line with the assumptions in this model that both tasks are energetically expensive and may be the same. Using information theory arguments only, one would expect that low effort tasks should elicit stronger responses, but that was not the case. From an energy point of view, a maximum $\mathcal{C}$, signals that the subsystem of neurons representing the outcomes are in a non-equilibrium state.

As in the previous section, the parameters are determined so that the distribution $p$ is a best estimate of the input distribution. For a fixed $q$, the parameters $\beta$ and $\lambda$ are determined numerically solely from the distribution of outcomes and initial conditions. The initial conditions $\beta = \lambda = 0, q = 1$ are used throughout this study which reflects an unbiased mental state of the DM regarding the outcomes of the gamble. We will be looking for solutions that maximize $\mathcal{C}$. Therefore, $\lambda/\beta$ will be negative. Among the possible solutions for different q, the solution with the highest q, $0<q\le 1$, is chosen (see Eq. \ref{entropydec}). Encoding the variance of the distribution of outcomes results in a normalization of the utility function of the values (or goods) of the gamble. 
It must be stressed that  even though \citet{Schwartenbeck2015} included variance in their analysis using the free energy method of \citet{Friston2009}, there is nothing in common with the treatment presented here which is motivated instead by energy considerations. Other work in decision making that was also inspired by thermodynamic is that of \citet{Ortega2013}, but no discussion of variance appeared in that work and their motivations were very different from the ones presented here. In either work, no attempt to discuss Birnbaum paradoxes was attempted. Our model, while motivated by the good-based model of \citet{Padoa2011}  of neurons, it is also developed to address paradoxical behaviors such those discussed extensively by \citet{Birnbaum2004b}.

\subsection{Applications: the case of two-gambles}

Consider two lotteries $A$ and $B$ similar to the ones treated earlier. Instead of considering each lottery separately, we now treat them simultaneously through the parameter $q$. Therefore, our lottery system is now represented in the mind by a joint probability distribution $p(A,B,\theta)$. We will assume that both distributions, $p(A)$ and $p(B)$ are independent but their effective entropy satisfies the q-entropy decomposition formula,
\begin{eqnarray}
H_q (A+B) = H_q(A) + H_q(B) + (1-q)H_q(A)H_q(B).
\label{entropydec}
\end{eqnarray}
The q-parameter is a dynamical parameter that microscopically may reflect the interaction of the neuronal populations encoding the gambles with the rest of the neuronal network \citep{Biro2005}. The $q=1$ case is where both gambles are completely independent, but it is also the case where their combined entropy is minimum for $0 < q \le 1$. This our motivation for choosing solutions with the largest q to minimize the combined entropy of gambles in a choice experiment.

As an application of this proposed model, we apply it to the Allais paradox, the  violation of stochastic dominance, and other new discovered paradoxes by \citet{Birnbaum2004b} that violate PT. When comparing two gambles, we will choose the one that has the higher $\mathcal{C}$, if $\mathcal{C}$ is well-behaved in the sense discussed above. Moreover, the q-parameter is the same if both gambles  are  compared simultaneously, but not necessarily the same across different pairs of gambles. For example, in Allais paradox we have two pairs of gambles that are presented to the DM at different times. From a physical point of view, the two pairs represent two different stimuli , and therefore it is expected that the q-parameters will be different too. 
But first, we take-up the example of the two lotteries with equal expected values and equal entropies.

\subsubsection{Gambles with equal expectation values and equal entropies}\label{evsection}

\begin{figure}[!htb]
	\includegraphics[width=7in,height=5in]{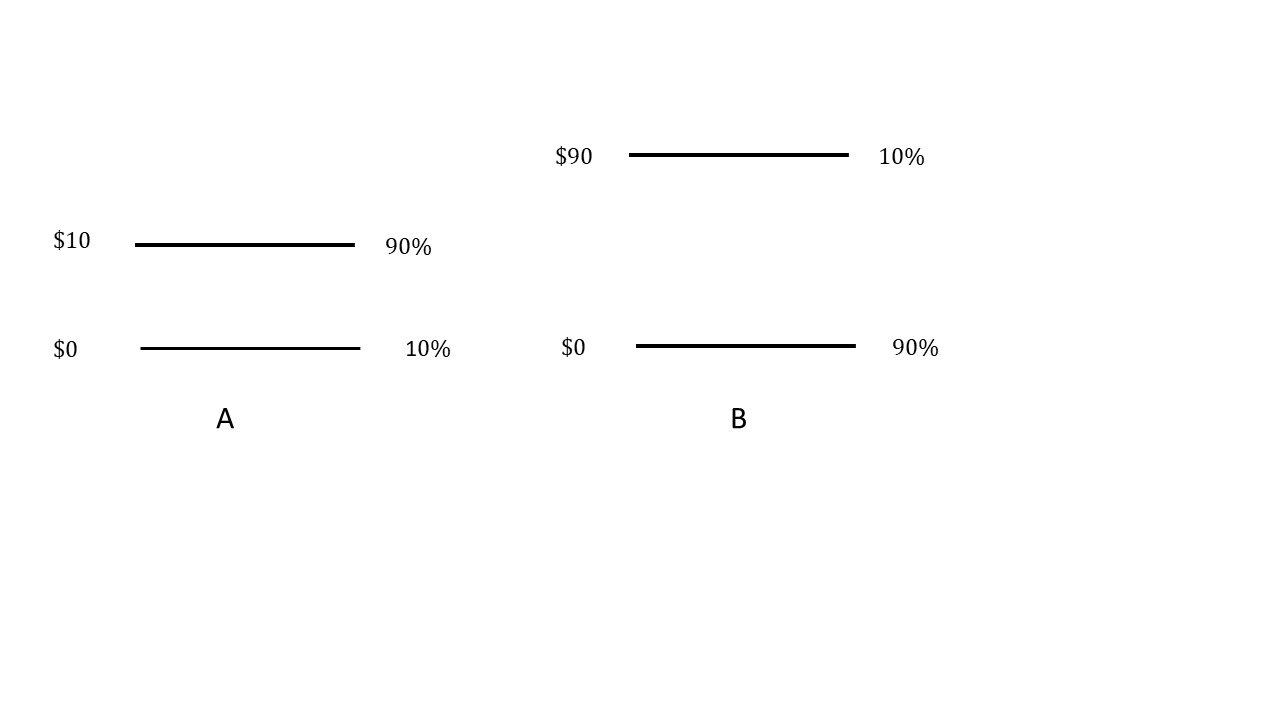}
	\caption{Gambles with equal expected value of $\$ 9$ and equal entropy, $H_q$ ($H_1 =-0.1 \ln 0.1 -0.9 \ln 0.9$). }
	\label{ev}
\end{figure}

As a first application of the entropic utility, we calculate the corresponding $\mathcal{C}$-values of both gambles that we discussed previously from a signal processing viewpoint. The gambles are represented pictorially in Fig. \ref{ev}. As we mentioned earlier, most people choose gamble A which is less risky, but PT fails to reproduce this result for  probabilities not so close to $100\%$, unless values are also distorted. Similarly, TAX does not predict that people prefer A unless values are also distorted. In this section, we will show that with both probabilities and values kept undistorted, we still can predict that people choose A over B. In a small sample of 19 people, we had only three people choosing B, that is about $85\%$ of people chose A. 

\begin{figure}[!htb]
	\includegraphics[width=7in,height=6in]{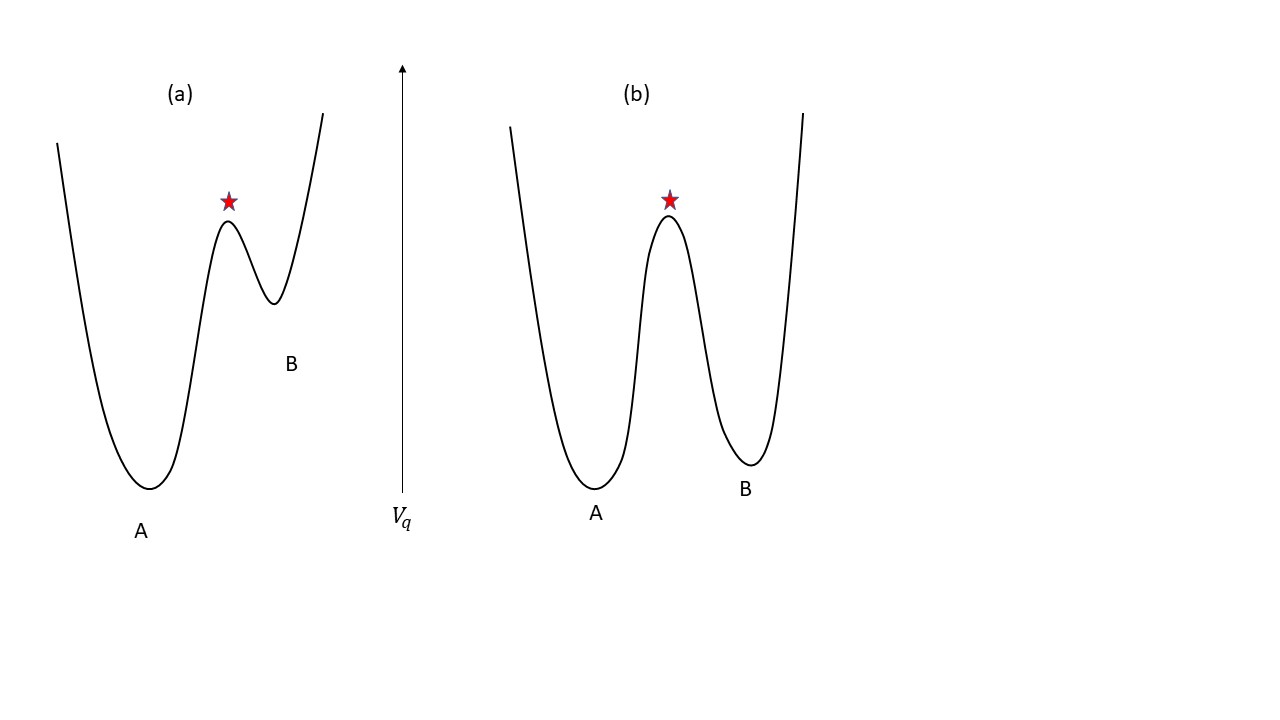}
	\caption{A qualitative shape of a dynamical theory of choice. A and B represent two possible choices. Making a decision between A and B should be more energetically favorable in case (a) than in case (b). The mental state represented by a "*" is the state we discuss in this work, i.e., just before the decision,  where both options are present.}
	\label{abpotential}
\end{figure}

Solving for $q$, we find several solutions. Among these solutions we choose the one where the difference between the $\mathcal{C}$-values is maximum. This difference can be associated with motivational momentum toward the chosen state and away from the suppressed state.
In Fig. \ref{abpotential}, we represent two cases where A is preferred to B from an energetic point of view, where in case (a) the choice of A is much easier to achieve than in case (b) which has a larger barrier between the two states. In this figure, we are sketching a potential for an interactive  theory, i.e., when both choices are entertained simultaneously in the mind of the DM. The star in the figure represents the metastable of the system right before a decision is made.  
\begin{table}[htbp]
	\centering
		\caption{Equal EV and equal non-extensive entropy gambles, Fig. \ref{ev}.}
	\resizebox{1.5\textwidth}{!}{\begin{minipage}{\textwidth}
			\label{tableev}
			\begin{tabular}{lllll}
				\hline
				& & q = 0.944 &  \\ \hline\hline
			  Gamble & 	&  A &  B  \\ \hline
				$\mathcal{C}$ &	& 0.0982 & -0.003  \\ \hline
				$\beta$	& & -22.79 & 30.75  \\ \hline
				$\lambda/\beta$ &	& -0.099 & -0.011 \\ \hline
			\end{tabular}
		\end{minipage}text}
\end{table}
Based on this criterion, the best values we found for this pair of gambles are shown in Table \ref{tableev}. Hence we reproduce the result that most people choose gamble A and shy away from gamble B. According to the relative signs of the $\beta$ terms in both gambles,  a DM who chooses gamble A over gamble B is entropy seeking in gamble A but entropy averse for entropy B. From an information theory point of view, this corresponds to increasing information about A while decreasing (or erasing) information about B \citep{Parrondo2015}. For both gambles,  a decrease in variance is energetically more favorable. This latter observation is always true in this analysis and follows from the sign of $\lambda/\beta$ which is always negative. A negative 'utility' can be interpreted that from an energy viewpoint, the cost of the processing of the information content exceeds that which is needed to maintain the state itself. However, it is more natural to interpret the difference as the direction of motivational momentum to drift toward one state and away from the other. The $\mathcal{C}$ value by itself is meaningless in this model, only differences are physical.

The current analysis that led us to the conclusion that A is preferred to B appears to be very different  from the signal analysis presented earlier. Both approaches however lead to the same conclusion in agreement with measurements but contradicts  Prospect Theory and the TAX model. The signal analysis discussion was at a more refined scale, where fluctuations caused by background noise was taken into account and probabilities were left linear and Boltzmann-Shannon entropy was used, but entropy is maximized in both cases. We expect a 
   microscopic treatment along the lines of \citet{Roxin2008}  to include aspects of both approaches, and will include higher order terms beyond variance, which is a  discussion beyond the scope of this work.

\subsubsection{Allais Paradox}

Next we apply our model to Allais paradox \citep{Allais1953}. In his work, Allais  argued that taking into account variance in addition to expected gain is essential to have a sound preference relation among gambles. To include information about variance, \citet{Meginniss1977} found that including an entropy function in addition to the expected utility function still provided a new utility function with properties mostly similar to EUT. The parameter $\beta$ in Eq. \ref{meginnis1} is a fitting parameter that is adjusted in such a way that Allais paradox can be explained on the basis of the new utility function. If the outcomes were normally distributed with variance $v$, then the entropy of outcomes is simply proportional to $\ln v$ \citep{Cover2006} and probably this was the motivation for Meginnis to consider entropy as a measure of variance. In our case, we are motivated on the basis of the space of outcomes embedded in a much larger space of states, that of the whole brain network, and therefore the constraints imposed on prospects  become very important to reproduce results in-line with behavior expectations. We follow the same analysis carried out in  section \ref{evsection}.

\begin{figure}[htb!]
	\includegraphics[height=5in,width=0.9\textwidth]{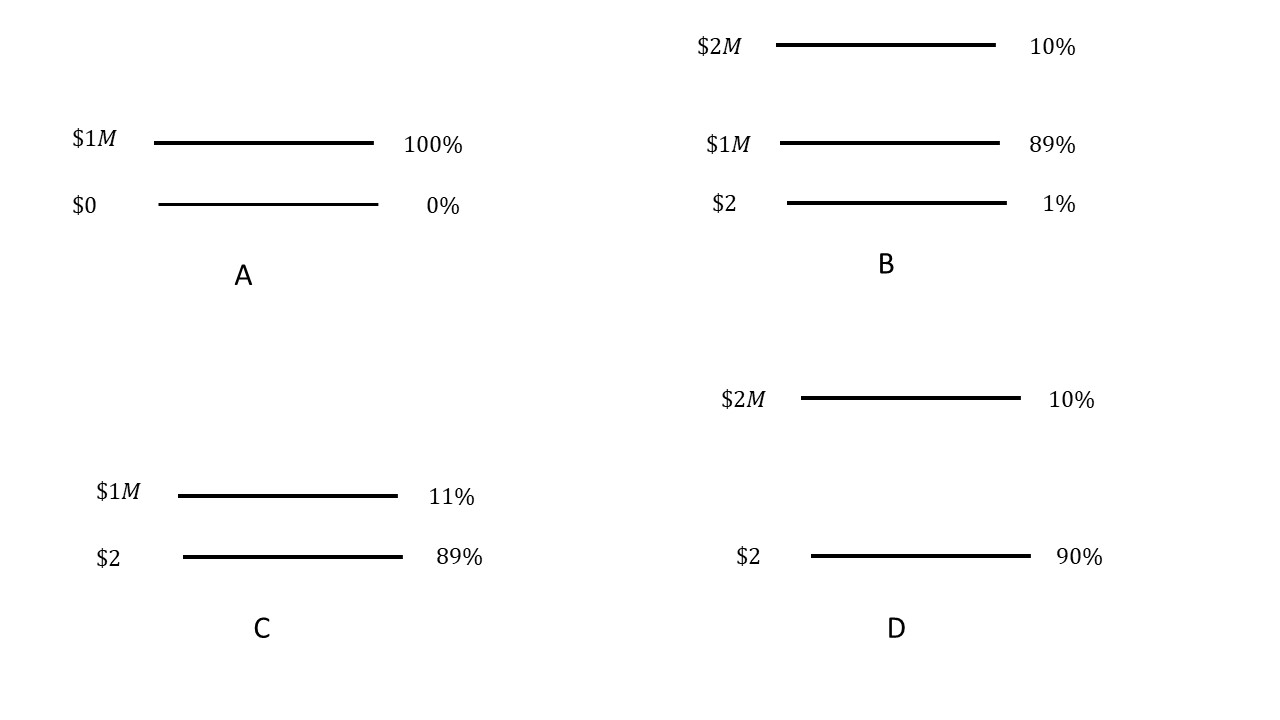}
	\caption{The Allais paradox discussed in \citet{Birnbaum2008}.}
	\label{allaisp1}
\end{figure}

The gambles that Allais proposed has the structure shown in Fig. \ref{allaisp1}. This particular example of four gambles has been tested extensively by \citet{Birnbaum2008}. 

Faced with a choice between gambles A and B, many people, 58\%,  would choose A rather than B. The only reason for not choosing B is the risky $1\%$ option of $\$2$, which is a very small risk, but people appear to favor sure outcomes in this case and show characteristics of a risk averse behavior. If faced with gambles C and D, many of the same people, $76\%$ would choose gamble D since the $\$ 2$ outcomes in both gambles are approximately the same. Therefore,  most people tend to show characteristics of a risk-seeking behavior in this case. Because of a $1\%$ outcome, we have a complete reversal in behavior which is odd for a linear theory such as EUT to satisfy. However, if nonlinearities in probabilities are incorporated, the behavior can be explained. The linearity in probabilities of EUT  makes common outcomes to be irrelevant to the preference relation and coalescing outcome values by adding their probabilities should not affect the preference relation either. The Allais paradox shows that people behavior cannot satisfy these seemingly 'rational' properties of EUT since a linear theory does not allow a reversal in preference, i.e., if A is preferred to B, it must be that C is preferred to D.

\citet{Birnbaum2008} attributes the failure of the Allais paradox to failure of coalescence of outcomes. From our point of view, it can be seen that coalescence modifies the variance of the distribution of outcomes but not the mean of the distribution which supports the idea of having variance as a quantity of relevance in decision making.

\begin{table}[htbp]
	\centering
	\caption{Allais paradox, Fig. \ref{allaisp1} in arbitrary units.}
	\resizebox{1.3\textwidth}{!}{\begin{minipage}{\textwidth}
			\label{tableallais}
			\begin{tabular}{lllll}
				\hline
				& & q = 0.713 &  \\ \hline\hline
				Gamble & 	&  A &  B  \\ \hline
				$\mathcal{C}$ &	& 100 & 56  \\ \hline
				$\beta$	& & -0.035 & -0.059  \\ \hline
				$\lambda/\beta$ &	& 0  & -4.51e-3 \\ \hline\hline
				&&&&\\
				&&  q = 0.378    &    \\ \hline\hline
				Gamble & 	&  D      &        C  \\ \hline
				$\mathcal{C}$ &	& 7.55e-3    & -8.52  \\ \hline
				$\beta$	& & 5.13    & -0.0033  \\ \hline
				$\lambda/\beta$ &	& -4.99e-3     & -7.27e-2 \\ \hline			
			\end{tabular}
		\end{minipage}text}
\end{table}

Since we are making a comparison between two gambles simultaneously, and we are looking for violations on an individual level, we expect the q-factor in both distributions to be the same. However, in the Allais paradox we are looking for consistency of decisions among two pairs. The q-factor is not necessarily the same in both pairs, but in the Allais paradox example, we could have chosen a q-factor that is common to both pairs and still find  the largest difference between gambles for the reasons we discussed above to be consistent with what most people choose. But this is not consistent with the physical picture we are advocating here. 
The q-factor can be understood as a residual interaction parameter of the underlying microscopic states of the gambles that are being compared only if both gambles are considered part of the metastable state at the same time. 

For the Allais paradox example in Fig. \ref{allaisp1}, the values we found that fits best the physical picture advocated here are displayed in Table \ref{tableallais}. The q-values found here give probabilities that have the usual inverse 'S'-shape as in Prospect Theory. 
 In both pairs of gambles, the DM is risk averse in the sense that negative changes in variance are more 'energetically' favorable. Note that our interpretation of the entropic utility function as an overall motivational momentum and that the extra parts as costs related to the information content is reasonable. For example, in the case of the certain lottery of $\$ 1M$, the information content is zero and hence there is no cost associated with it. However, differences are what is physically relevant here, since they give the direction of motivational momentum between the states:    $ 44 = \mathcal{C}_A -\mathcal{C}_ B > \mathcal{C}_D -\mathcal{C}_ C = 8.5$.  According to this inequality, if reaction times are measured, people will tend to respond faster when choosing $A$ compared to when choosing $D$.
The negative sign of the parameter $\beta$ implies that the higher value states are being more weighted in the decision than lower value states.


\subsubsection{Stochastic Dominance}

 Given two gambles A and B, if the
probability of winning x or more in gamble A is greater
than or equal to the probability of winning x or more in
gamble B for all values of x, and if this probability is
strictly higher for at least one value of x, we say that A
stochastically dominates B. Therefore mathematically, stochastic dominance will involve calculating a cumulative probability density of a distribution.
In short, stochastic dominance expresses the apparently rational fact that more is better. This is the reason why stochastic dominance violations were not easily accepted.  \citet{Birnbaum2004b} extensively studied this violation under many different frameworks, and obtained strong evidence that violation of stochastic dominance is a very common behavior among DMs.

The violation of stochastic dominance is accounted for by the old version of Prospect Theory \citep{Kahneman1979}, the TAX model \citep{Birnbaum2008}, and the Attention model \citep{Johnson2016} for varying reasons. We show here that the  theory presented above also allows for  stochastic dominance to be violated. The pair of gambles we have chosen to discuss for this violation is from \citet{Birnbaum2004b}:
\begin{eqnarray*}
	J & = & (\$ 12, 0.10; \,\, \$ 90, 0.05; \,\, \$ 96, 0.85)\\
	I & = & (\$ 12, 0.05; \, \,\$ 14, 0.05; \, \, \$ 96, 0.90).
\end{eqnarray*}

Gamble I stochastically dominates J because the
probabilities of winning $\$ 14$ or more and of $\$ 96$ or more
are greater in gamble I than J, and the probabilities of
winning $\$ 12$ or more and of $\$ 90$ or more are the same in
both gambles. This is easily seen if both gambles are rewritten in the following way by {\it splitting} the outcomes:
\begin{eqnarray*}
	J^* & = & (\$12, 0.05;\,\, \$12, 0.05; \,\, \$90, 0.05; \,\, \$96, 0.85), \\
	I^* & = & (\$12, 0.05; \,\, \$14, 0.05; \,\, \$96, 0.05; \,\, \$96, 0.85).
\end{eqnarray*}
 If the two lotteries are compared term by term, it is clear that according to the laws of probability, lottery $I$ should be preferred to lottery $J$. But this contradicts what most people choose to do. However, from a heuristic point of view, it is easy to see why most people would prefer to play lottery $J$ if they want to maximize their chances of winning bigger amount of money. People choose J over I presumably because
 J has two ways to produce a ‘very good’
 outcome but I has only one way (as if the
 outcomes are equally likely even though they
 are not) \citep{Birnbaum1996}. This latter argument by itself indirectly shows the relevance of variance in the decision process, since both $I^*$ and $J^*$ have the same entropy \citep{Schwartenbeck2015}, but in the initial set-up, J has higher entropy than that of I. Clearly linear operations on the gambles, as described here, is not a step that the human brain uses to compute the effective value of each gamble, and variance brings the minimal nonlinearity needed in the computations to make them align with human behavior. According to \citet{Birnbaum2008}'s TAX model, people fail to detect stochastic dominance because the coalescence or splitting of branches with equal outcomes is not a valid step. From the point of view of our model, coalescence is clearly violated because of the nonlinearities in the probabilities and in the outcomes.
\begin{figure}[h!]
	\includegraphics[height=5in,width=1.2\textwidth]{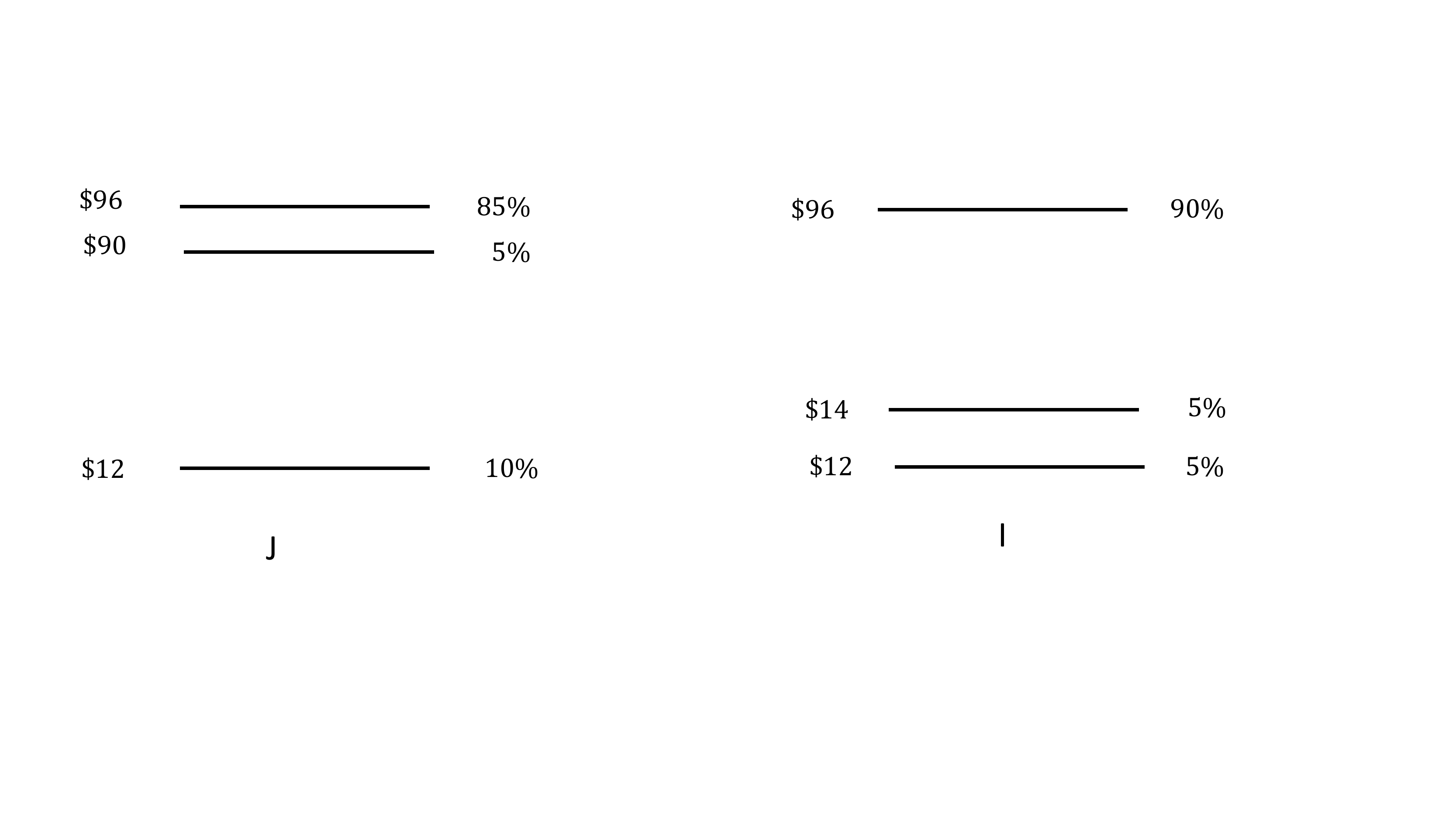}
	\caption{A lottery that violates stochastic dominance \citep{Birnbaum2004b}.}
	\label{stocIJ}
\end{figure}

For the pair of gambles J and I displayed in Fig. \ref{stocIJ}, \citet{Birnbaum2004b} conducted 16 different studies over the years. The violation of stochastic dominance was significant in all of them. The theory we proposed here also shows that violations of stochastic violation is valid for all $0.33 \le q \le 1$. The q-value, $q = 0.801$ we have chosen is the one that also works for a transparent stochastic dominance example related to the $I-J$ pair, but this does not have to be the case as we discussed above. The pair of gambles $J_1$ and $J$ \citep{Birnbaum2004b},
\begin{eqnarray*}
J_1 & = & \left( \$ 12, 0.10; \,\, \$96, 0.90 \right) 
\end{eqnarray*} 
are easily seen to satisfy a dominance relation where $J_1$ stochastically dominates $J$. The chosen q-value for the probability distributions respects stochastic dominance for the pair $J_1 - J$ and violates stochastic dominance for $I-J$.
\begin{table}[htbp]
	\centering
	\caption{Violation of stochastic dominance \citep{Birnbaum2004b}, Fig. \ref{stocIJ}.}
	\resizebox{1.3\textwidth}{!}{\begin{minipage}{\textwidth}
			\label{tablestoc}
			\begin{tabular}{lllll}
				\hline
				& & q = 0.801 & &  \\ \hline\hline
				& & & & \\
				Gamble & 	&  J & \,\,\,\, \, &  I  \\ \hline
				$\mathcal{C}$ &	& 6.5  &      & -266.1  \\ \hline
				$\beta$	& & 0.554  &          & 0.0098  \\ \hline
				$\lambda/\beta$ &	& -9.68e-3 &\,\,\, \,\, & -3.86e-2\\ \hline
			\end{tabular}
		\end{minipage}text}
\end{table}

The parameters derived for the $I-J$ pair are given in Table \ref{tablestoc}. Again using our analogy with energy, the relative costs associated with the information content  of gamble I are much more than that of gamble J. This is reflected in the negative sign of $\mathcal{C}$. According to our model, the reason people prefer J to I has to do more with the cost of information associated with the gamble rather than its utility, or differently, people are less motivated to drift to gamble I. Since in our model we are assuming a linear relationship between utility of outcomes and energy, the opposite of $\mathcal{C}$, $-\mathcal{C}$, should be interpreted from the point of view of the rest of the neuronal network as the excess energy needed to maintain the encoding of the gamble in the brain. It is this excess of energy that is minimized by the (computational part of the) brain. So it is important to remember that in classic descriptive theories, the utility is synonymous with the DM, but here utility is just the energy needed to represent the gamble in one part of the cortex. 

Finally, we would like to make further comments on the choice of the parameter $q$. According to the physical picture advocated here, the $q$ parameter should reflect the interaction of the two lotteries that are being considered and no other. For each pair of gambles, we need just to find $q$ that gives the largest difference between the two gambles. In this case $q=1$ should have been chosen for the $IJ$ gamble. The parameters are still approximately the same and our conclusions won't be affected. For $q = 0.801$, $\Delta \mathcal{C}_{J-I} = 273$, while for $q = 1$, $\Delta \mathcal{C}_{J-I} = 275$. In general however, we should choose $q$ independently for each pair of lotteries. This is what we would do for the remaining paradoxes.


\subsubsection{Birnbaum Paradoxes}

In Savage's subjective utility theory, the {\it "sure-thing"} principle is an important assumption in the theory. The principle states that if a DM prefers A to B given that event C occurred or not-C occurred, then the DM should still prefer A to B even if the DM does not know anything about C \citep{Savage1954}. This principle is now only of general historical interest since people do not obey this principle even in it weakest form that of branch independence which has been shown to be violated \citep{Birnbaum1996}. Branch independence implies that outcomes or branches with equal outcomes should not matter when comparing two gambles.  Birnbaum paradoxes are a manifestation of violations of branch independence and coalescence \citep{Birnbaum1998}.

To further show that coalescence is at the root of these violations, \citet{Birnbaum2004b} studied  many  gambles that he expected to violate Prospect Theory. In this paper, we will address the violations of upper and lower cumulative independence (Table 3 in \citet{Birnbaum2004b}). For outcomes $0<z<x'<x<y<y'<z'$, and probabilities $p, q, $ and $r$,  upper cumulative independence is expressed as follows:
\begin{eqnarray}
 & S' =(x,p;y,q;z',r)  <  R'=(x',p;y',q;z',r) \\
 \Rightarrow & S'''=(x,p+q;y',r) < R'''=(x',p;y',q+r). \nonumber
\end{eqnarray}
For lower cumulative independence, we have
\begin{eqnarray}
& S=(z,r;x,p;y,q) > R =(z,r;x',p;y',q) \\
  \Rightarrow & S''=(x',r;y,p+q) > R''=(x',r+p;y',q). \nonumber
\end{eqnarray}
All theories with weights that depend on a cumulative  distribution of outcomes , including Prospect Theory and rank dependent utilities, satisfy these properties. These two properties are not dependent on the inverse S-shape of the weight function, and the reader is referred to \citet{Birnbaum1998} for more details about how these two properties are derived.  Therefore,  to show the robustness of the violations of both of these properties, Birnbaum employed different forms of the stimuli to minimize any framing effects.  
His TAX model is able to reproduce these new violations. In the following, we show that these new paradoxes can be also explained within our model.

\paragraph{Upper cumulative independence:}

\begin{figure}[H]
	\includegraphics[height=5in,width=0.95\textwidth]{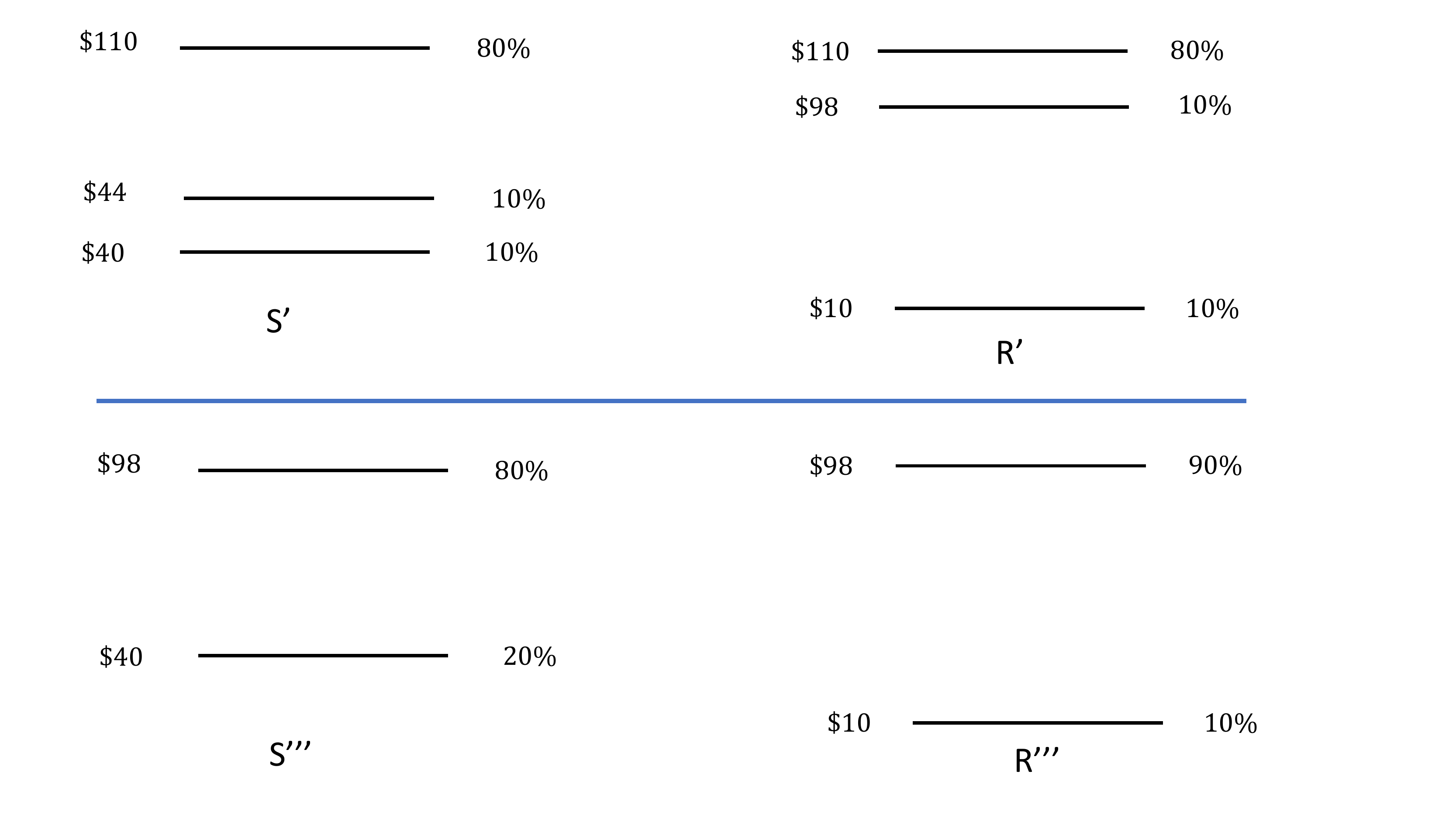}
	\caption{Upper cumulative independence: pair of gambles no. 10 and no. 9 in \citet{Birnbaum2004b}.}
	\label{uppercum}
\end{figure}

Upper cumulative independence (UCI) were predicted by TAX but violated by PT. An example of UCI is displayed in Fig. \ref{uppercum}. For the gambles, we will use the same notation as that of \citet{Birnbaum2004b}. The violations of UCI are due to failure of coalescence and branch independence. According to PT, if a person prefers $R'$ over $S'$, then that person should also prefer $R'''$ over $S'''$. Here $S'''$ is obtained from $S'$ by coalescing the lower two outputs, while $R'''$ is obtained from $R'$ by coalescing the upper outcomes. In the 12 studies reported by Birnbaum testing UCI, he found that about $69\%$ of people preferred $R'$ to $S'$ while about $63\%$ of people switched preference in the second pair of gambles, i.e., most chose $S'''$. 

The parameters that are consistent with a violation of UCL are shown in Table \ref{tableUCI}. For each pair of gambles, the q-value corresponds to the maximum difference in utilities and are in the expected preference order as chosen by people. In comparing gambles $R'$ and $S'$, the difference is 
$\Delta\mathcal{C}_{R'S'} = 36.72$, while that between the gambles $S'''$ and $R'''$ is 
$\Delta\mathcal{C}_{S'''R'''} = 6.37 \times 10^3$. Therefore, there is a reversal in preferences.

\begin{table}[H]
	\centering
	\caption{Violation of upper cumulative independence of pair of gambles (10,9) in Fig. \ref{uppercum}, \citep{Birnbaum2004b}.}
	\resizebox{1.3\textwidth}{!}{\begin{minipage}{\textwidth}
			\label{tableUCI}
			\begin{tabular}{lllll}
				\hline\hline
				& & q = 1  & & \\ \hline\hline
				&&&&  \\
				Gamble & 	&  $R^\prime$ &  & $S^\prime$  \\ \hline
				$\mathcal{C}$ &	& -3.23  & & -39.95  \\ \hline
				$\beta$	& & 0.19  &  &  0.038  \\ \hline
				$\lambda/\beta$ &	& -9.47e-3 & & -1.21e-2 \\ \hline\hline
				&&&& \\
				\hline\hline
					& & q = 0.368  & & \\ \hline\hline
					&&&& \\
				Gamble & 	&  $S^{\prime\prime\prime}$ & & $R^{\prime\prime\prime}$  \\ \hline
				$\mathcal{C}$ &	& 27.8  &  & -6. 34 $10^3$  \\ \hline
				$\beta$	& & 3.70  &  &  3.3e-4  \\ \hline
				$\lambda/\beta$ &	& -7.23e-3 &  &  -0.70  \\ \hline
			\end{tabular}
		\end{minipage}text}
\end{table}
\begin{figure}[H]
	\includegraphics[height=5in,width=0.95\textwidth]{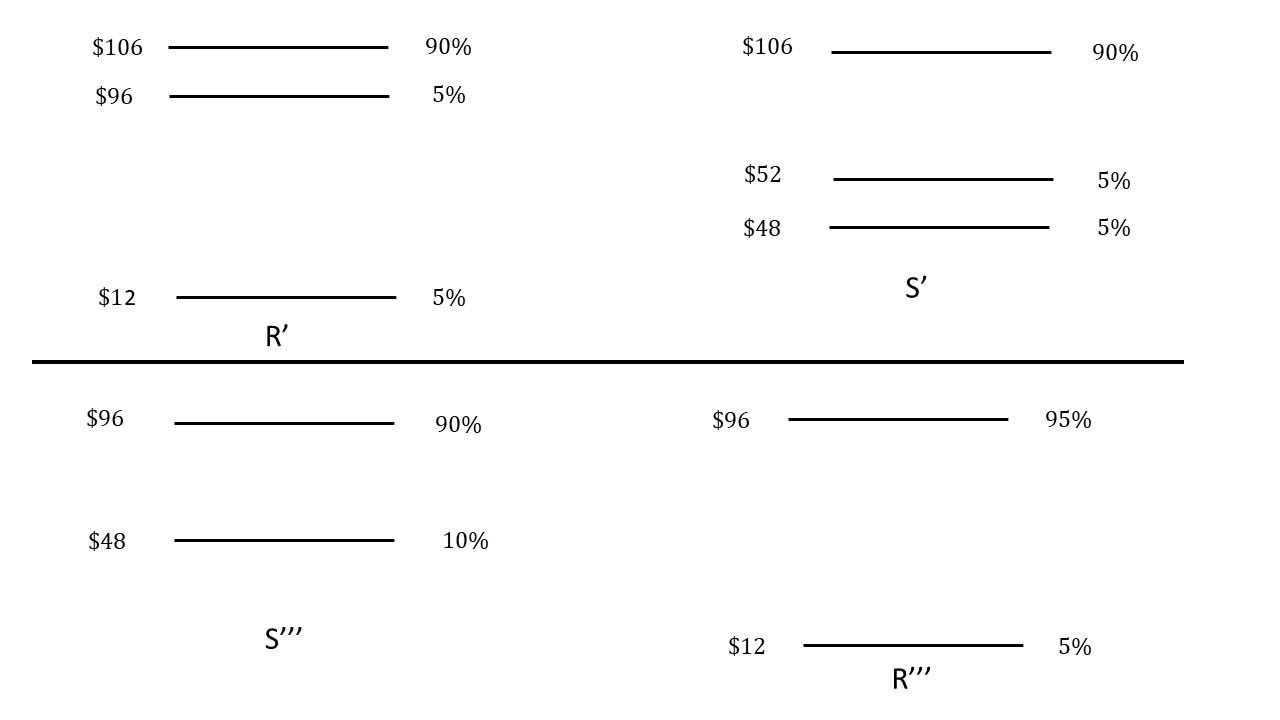}
	\caption{Upper cumulative independence: pair of gambles no. 12 and no. 14 in \citet{Birnbaum2004b}.}
	\label{uppercum2}
\end{figure}
The second pair of gambles discussed by Birnbaum is shown in Fig. \ref{uppercum2}. The parameters corresponding to this pair of gambles are given in Table \ref{tableUCI2}. According to the experimental measurements, about $52\%$ of people chose $R'$ over $S'$, while about $71\%$ of people chose $S'''$ over $R'''$, which is a violation of UCI. The much higher percentage in the $S''' > R'''$ is reflected in the magnitude of differences of entropic utilities,
$\Delta\mathcal{C}_{R'S'} = 9.14$, for the first pair and 
$\Delta\mathcal{C}_{S'''R'''} = 7.91 \times 10^2$ for the second pair of lotteries. 

\begin{table}[H]
	\centering
	\caption{Violation of upper cumulative independence of pair of gambles (12,14) in Fig. \ref{uppercum2},  \citep{Birnbaum2004b}.}
	\resizebox{1.3\textwidth}{!}{\begin{minipage}{\textwidth}
			\label{tableUCI2}
			\begin{tabular}{lllll}
				\hline\hline
				& & q = 1  & & \\ \hline\hline
				&&&&  \\
				Gamble & 	&  $R^\prime$ &  & $S^\prime$  \\ \hline
				$\mathcal{C}$ &	& 1.65  & & -7.5  \\ \hline
				$\beta$	& & 0.33  &  &  0.092  \\ \hline
				$\lambda/\beta$ &	& -9.40e-3  & & -1.0e-2 \\ \hline\hline
				&&&& \\
				\hline\hline
				& & q = 0.737  & & \\ \hline\hline
				&&&& \\
				Gamble & 	&  $S^{\prime\prime\prime}$ & & $R^{\prime\prime\prime}$  \\ \hline
				$\mathcal{C}$ &	& 32.33  &  & -7.58e+2  \\ \hline
				$\beta$	& & -8.29  &  &  3.13e-3  \\ \hline
				$\lambda/\beta$ &	& -6.88e-3 &  &  -9.30e-2  \\ \hline
			\end{tabular}
		\end{minipage}text}
\end{table}

\paragraph{Lower cumulative independence:}

The pair of gambles we use to show violations of lower cumulative independence (LCI) is from \citet{Birnbaum2008}. The gambles were studied under 12 different frames \citep{Birnbaum2004b}. According to LCI, if a person chooses $S$ over $R$, then that person should also choose $S''$ over $R''$.  Prospect Theory satisfies LCI. 

\begin{figure}[H]
	\includegraphics[height=5in,width=0.95\textwidth]{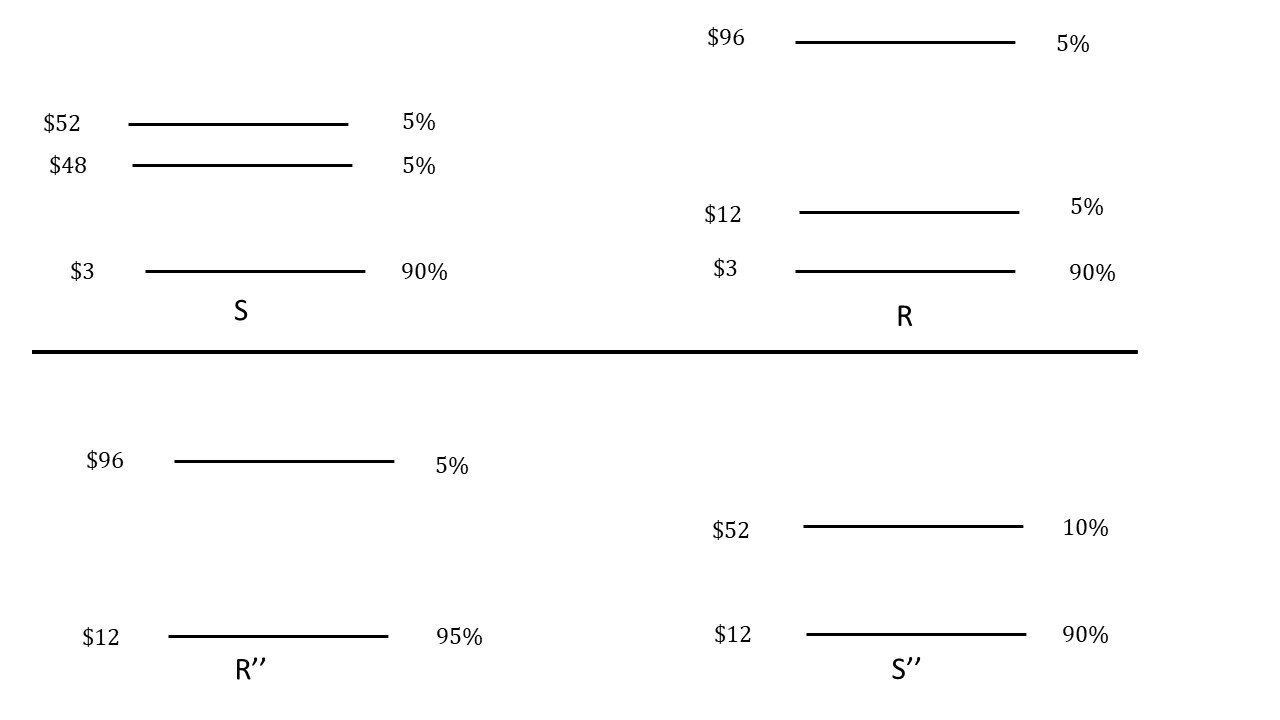}
	\caption{Lower cumulative independence: pair of gambles no. 17 and no. 20 in \citet{Birnbaum2004b}.}
	\label{lowercum2}
\end{figure}

The pair of gambles are shown in Fig. \ref{lowercum2} with the corresponding parameters in Table \ref{tableLCI2}. According to the measurements, about $51\%$ of people prefer $S$ over $R$, while about $65\%$ prefer $R''$ over $S''$ which is a violation of LCI.  The parameters determined from our theory for this pair of gambles are given in Table \ref{tableUCI2}. The effective motivational effort, $\mathcal{C}$,  is larger for the $R''S''$ pair than for the $RS$ pair,  $\Delta \mathcal{C}_{SR} =  0.53 \,\, <  \,\,  \Delta \mathcal{C}_{R"S"} = 0.92$. This is indirectly reflected in the percentage numbers quoted above, which imply that it is cognitively easier for the process $R'' > S''$ to be processed than that of $S > R$. Note that according to our theory, we cannot conclude for certain that $\mathcal{C}_{R"S} = 6.96$ is larger than $\mathcal{C}_{S''R} = 6.54$ since all the parameters will be different, but what we can conclude is that the reaction time to choose $S$ should be slower than that of choosing $R''$.  Only a fully dynamical theory, with higher-order terms in values, can account for these differences in transition rates. In the next section, we propose some ideas on how to approach this problem using the Maxwell 'demon' analogy \citep{Parrondo2015}.

\begin{table}[H]
	\centering
	\caption{Violation of lower cumulative independence of gambles (17,20) in  Fig. \ref{lowercum2}, \citep{Birnbaum2004b}.}
	\resizebox{1.3\textwidth}{!}{\begin{minipage}{\textwidth}
			\label{tableLCI2}
			\begin{tabular}{lllll}
				\hline\hline
				& & q = 0.506  & & \\ \hline\hline
				&&&&  \\
				Gamble & 	&  $S$ &  & $R$  \\ \hline
				$\mathcal{C}$ &	& 3.74  & & 3.21  \\ \hline
				$\beta$	& & 0.128  &  &  0.363  \\ \hline
				$\lambda/\beta$ &	& -1.0e-2   & & -1.0e-2   \\ \hline\hline
				&&&& \\
				\hline\hline
				& & q = 0.974  & & \\ \hline\hline
				Gamble & 	&  $R^{\prime\prime}$ & & $S^{\prime\prime}$  \\ \hline
				$\mathcal{C}$ &	& 10.7  &  & 9.75   \\ \hline
				$\beta$	& &  77.76     &  &  72. 6 \\ \hline
				$\lambda/\beta$ &	& -9.26e-3 &  &   -1.56e-2  \\ \hline
			\end{tabular}
		\end{minipage}text}
\end{table}

\vspace{2cm}

\section{An Interactive Model for Choice with Memory}

Why it is not enough to only compare lotteries by calculation of expected utility separately? While this approach appears to work in many cases, it does not work when decoy effects or framing effects are included. For example, it is known that questions posed in terms of losses instead of gains may reverse the choice made by the DM. Similarly, addition of irrelevant options can affect the choice \citep{Vlaev2011}. At the neuronal level, it has also been detected in single trials that changes in preferences can occur right before the commitment to a decision \citep{Kiani2014}. Moreover, we know that inhibiting neurons play a role in the dynamics leading to a decision, but a full picture of the mechanisms of choice are still not known \citep{Hayden2018,Yoo2018}. Therefore, extensions of choice mechanisms that include interaction among the different options is desirable.

So far, we have been able to argue based on a physical picture and plausible physical arguments to present an account of behavioral decision making under risk. The results we were able to obtain were very encouraging, but this is still far from a physically sound theory that can bridge the gap between behavioral and neurobiological processes. Another improvement of the treatment presented above is to propose a scheme where interactions between gambles that are being compared simultaneously are included in the analysis. To achieve this, we need to account for other processes that take part in a decision process such as attention and memory, and hence extra parameters will need to be included.  

The need to incorporate attention in models that explain cognitive tasks is well documented \citep{Pashler1998}. For example, 
 eye-tracking analysis of the dynamics of decision making shows that attention increases with increasing probability and value \citep{Fiedler2012,Johnson2016}.  This attention process will be regulated by another population of neurons in addition to the populations considered above that encoded only the gambles. However, to include attention in our model a memory element is needed to be part of the model. 
 This memory works as a register of the transitions that are unconsciously or consciously being made between the various branches of the two gambles to arrive to a decision. The actual decision step was only included in an indirect manner in the above discussion. To improve on this picture, we use the Maxwell demon analogy in nonequilibrium statistical mechanics \citep{Mandal2012}. Just to remind the reader, Maxwell demon is an "intelligent" being suggested by Maxwell that circumvented the second law of thermodynamics \citep{Maxwell1891}. The demon used information about every particle of a gas in equilibrium to decrease entropy which is a violation of the second law of thermodynamics. In a classic paper, \citet{Szilard1929} showed that when the memory of the demon is included part of the system, entropy can no longer decrease. This work established for the first time the physical nature of information. The reader is referred to the review by \citet{Leff2002} for more details.  For our purposes, the demon will be the equivalent of the working memory  of the DM. The Maxwell demon lowers the entropy of the subsystem by directing the flow of particles in the gas following a control protocol. Similarly, the WM will be involved in directing the 'flow' of attention between the states (branches) of the different gambles. In our decision problem, there are transitions of attention within each gamble and  across gambles. Both of these types of transitions will be manipulated by the demon using a Master equation approach with probability transition rates determined from the nonextensive entropy approach. One of the  candidates for the Maxwell demon in the brain
could be the thalamic reticular nucleus (TRN). The TRN is known to control traffic between the cortex and thalamus \citep{Lam2011}. It has also been shown that RTN plays a role in regulating attention between competing visual and auditory stimuli \citep{Wimmer2015}. This means that TRN plays some role in deciding which information is more relevant to the goal set in the brain.

\subsection{Master equation approach to attention}

From an energetic viewpoint, consider two lotteries L and R, the population of neurons representing each gamble will form an instantaneous equilibrium, and during the deliberation phase of the decision making, the decision maker is assumed to explore all possible states by a feedback process that also sustains the signal in the absence of the stimulus. The feedback loop process is similar to the rehearsal component of the \citet{Atkinson1968} model for short-term memory. For the decision to be made, both gambles must remain active in a short-term buffer in prefrontal cortex such as in OFC and may be other regions of the cortex. In our memory, both implicit and explicit attentions are involved, and may span different time scales depending on the relative values of the parameters that we determined earlier in the entropic utility approach.

\begin{figure}[H]
	\includegraphics[width=8in,height=4in]{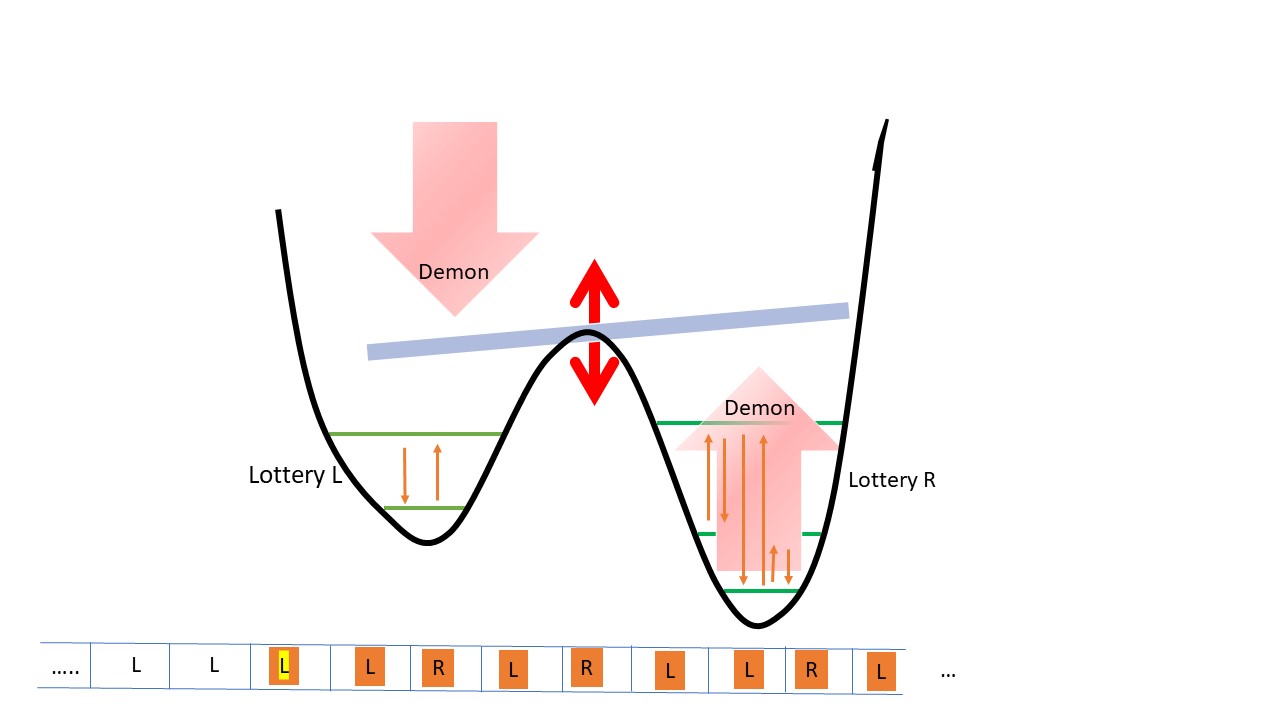}
	\caption{The system is made of two subsystems, $L$ and $R$, that represent the lotteries. The tape represents the memory of the mind as it juggles between the two subsystems. The distribution of L's and R's in the outgoing tape represents the distribution of choices made by people and represented by variable $\theta = 0,1$. Here it is assumed that people are first presented with the lottery at the left. So initially, people register lottery L in their memory, and then afterwards decide to stay in L or move attention to R. Moving attention away from L is equivalent to erasing information about L. In this particular example, a single time step is assumed to start always from the L state.}
		\label{doublewell}
\end{figure}

In the deliberation phase, shifting attention among  all possible states will be part of the dynamical process  that will upset the local equilibrium and drive the system toward the chosen state. The analogy with a chemical reaction in the presence of an enzyme is very close to the picture we are adopting her. Schematically, the dynamic of decision is represented in Fig. \ref{doublewell}. Each well represents a lottery with transitions represented by thin arrows, while fat arrows represent transitions between states. The demon is represented here by a tape that represents bias or an energy-feedback loop that facilitates the decision to proceed in the least resistant path in the neuronal network and keeps track of the transitions made between L and R. We will not treat the whole dynamic of the system in a self-consistent way, but we will introduce the minimum ingredients needed to account for memory of the demon. The discussion of entropy manipulation by WM will not be discussed here, but may give a better picture of how entropy generation is transported across the different regions of the brain.  Instead, we will slightly extend the previous discussion and focus on accounting for the overall probabilities of choosing either gamble given the outcomes and weights of the different branches in a pair of gambles.

In the sensorimotor region, different action options are represented in parallel \citep{Gold2000}.  Hence, it is reasonable to represent different options with different states in parallel. Moreover as we discussed above, there is some evidence in a two-choice reaching task that neural activity in the primary cortex (M1) and dorsal premotor (PMd) regions does not integrate information, it just tracks the state of the sensory state \citep{Thura2014}. In value-based decision tasks, we have less evidence that evidence accumulation is the main pathway for reaching a decision. Given these uncertainties, a master equation approach to the probability distribution of the 'macroscopic' states of gambles is proposed here \citep{Gardiner1985}.

For a Markovian system, the rate of change of probabilities is given by
\begin{eqnarray}
\frac{d p_i}{d t} = \sum_j p_j W_{ji} - p_i W_{ij}
\label{mastereq}
\end{eqnarray}
where $W_{ij}$ is the transition probability rate from the $i$-th state to the $j$-th state. The transition probabilities within each gamble are determined from the probability distributions found earlier, while the transition between gambles are caused by the demon, which are assumed constant for simplicity. These transition rates give rise to a flow  of entropy  in or out of the coarse-grained subsystem describing the gambles to the rest of the brain. The picture we adopt here is very similar to the one used in systems with thermodynamic and information processes \citep{Parrondo2015}. Therefore, the introduction of the demon will drive the subsystem to a non-equilibrium state. The coefficients $W_{ij}$ will be interpreted in terms of attention \citep{Nunez2017}.

To keep our model manageable, we will assume local detailed balance for transitions within a gamble. This implies that the   the transition probability rate from state, $W_{ij}$( $i \rightarrow j$) satisfy the following relation between any pair of outcomes or states:
\begin{eqnarray}
W_{ij} p_i^{eq}   & = & W_{ji} p_j^{eq}.
\end{eqnarray}
Fortunately for us, this non-equilibrium state can achieve a steady state which makes us avoid introducing time in the decision problem explicitly. At the individual level, this is a poor approximation, since we are assuming that the DM is taking infinitely many time steps to reach a decision. However, by ergodicity this final distribution is the same as that achieved by an 
 infinitely large sample of individual brains with arbitrary initial states. Therefore, the reported distribution of choices between two lotteries is closely related to the equilibrium distribution that we calculate in each choice task.

In the nonextensive entropy model, this leads to the following expression
\begin{eqnarray}
\frac{W_{ij}}{W_{ji}} & = & \left[\frac{1-(1-q)\sum_\nu \lambda_\nu^* (O_{\nu,i} - u_\nu)}{1-(1-q)\sum_\nu \lambda_\nu^* (O_{\nu,j} - u_\nu)}  \right]^{\frac{1}{1-q}}
\end{eqnarray}
where the $\lambda_\nu^*$ are the normalized Lagrange multipliers and $\sum_i O_{\nu,i} = u_\nu$ is the $\nu$th constraint on the subsystem.  
We choose these transition rates as
\begin{eqnarray}
W_{ij}  & = & \Omega (\beta,\lambda) \left[\frac{1-(1-q)\sum_\nu \lambda_\nu^* (O_{\nu,i} - u_\nu)}{1-(1-q)\sum_\nu \lambda_\nu^* (O_{\nu,j} - u_\nu)}  \right]^{\frac{1}{2}\frac{1}{1-q}},
\label{wij}
\end{eqnarray}
where $\Omega$ is the transition rate of attention within a single lottery. If transition state theory is any guidance,  we expect to have $\Omega$ of the same order as  $1/\beta$. 
 Hence both parameters $\Omega$ and $\omega$ represent the frequency of shifting attention among the outcomes of a single lottery and between lotteries.  It may be possible that an eye fixation analysis can be used to estimate some of these parameters \citep{Rosen1976}. In addition to these two parameters, we also need a distribution in working memory of the two gambles. This distribution will eventually represent the belief or information gained after the deliberation (i.e, feedback) process. Based on this distribution, the working memory will erase any information associated with the discarded lottery (i.e., decrease in entropy), and 'focus' attention on the chosen lottery. This information is then relayed to a new population of neurons in the OFC that encode the chosen state \citep{Padoa2014,Rustichini2017}. Padoa {\it et al.} were able to  identify a different set of neurons that encoded the chosen good (juice). Our model is based on his findings, and chosen gamble is represented in the working memory (WM), or demon. This is best illustrated by treating various examples.    
 
\subsection{Applications II}

\subsubsection{Stochastic Dominance}
\begin{figure}[H]
	\includegraphics[width=6in,height=3.5in]{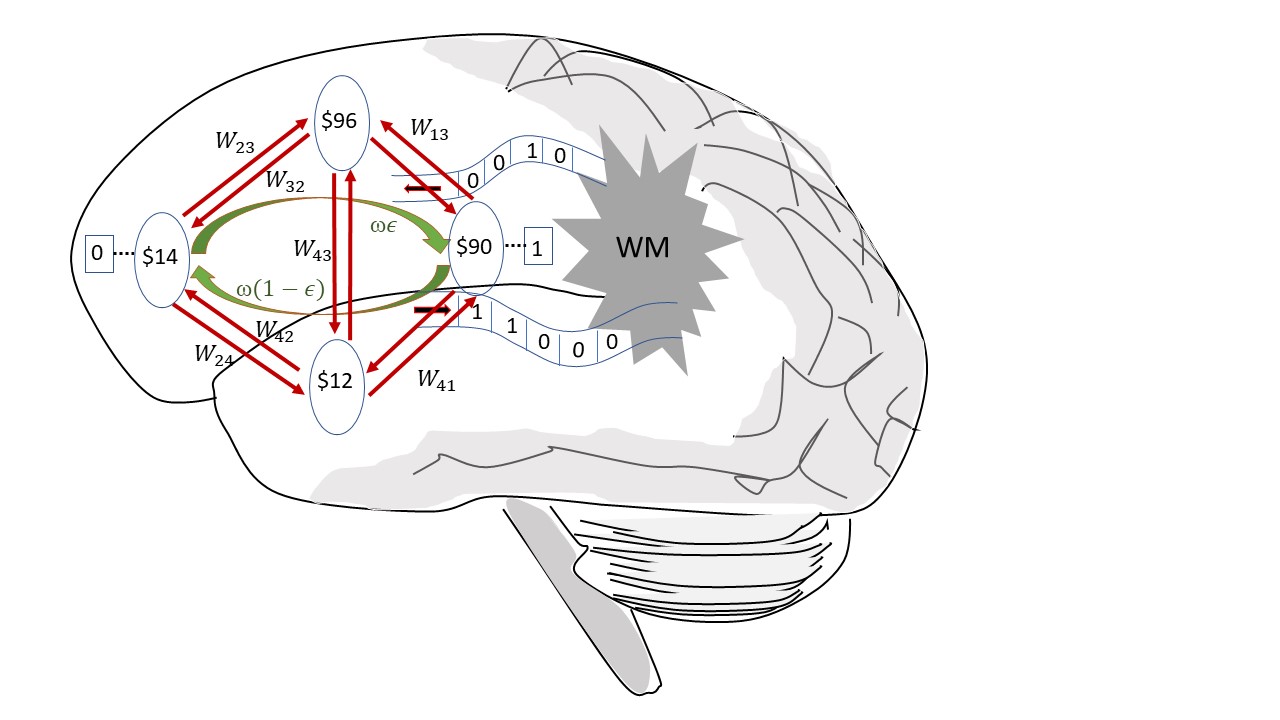}
	\caption{Stochastic dominance circuit: Choice between gamble $J = (\$ 96, \, 0.85, \$ 90, \, 0.05, \$ 12, \,0.10)$ and gamble $I = ( \$ 96,\, 0.90, \$ 14,\, 0.05, \$ 12, \,0.05)$. The gamble J is stochastically dominated by gamble I \citep{Birnbaum2004b}. The straight arrows (red online) represent transitions between the various gains. The green (online) curly arrows represent transitions between the two options I and J. The working memory (WM) stores and controls transitions between I and J.    }
	\label{stochdom}
\end{figure}

The first example we treat is that of stochastic dominance discussed earlier. Since we are addressing switching of attention among the various possible outcomes of both lotteries, we need first to set-up the possible paths along which these transitions of attention can be activated. The working memory will play the role of a traffic controller between various states that are usually the focus of attention. The circuit that corresponds to the two lotteries in Fig. \ref{stocIJ} is shown in  Fig. \ref{stochdom}, and is represented in an abstract value-based space. In addition to the probability transition rates, $W_{ij}$, that we already discussed, there is a new element in this network that corresponds to the working memory that is responsible for new attentional guidance between the various options available to the DM. Similar processes were observed in the visual system where the WM modulates competitive interactions between items in the visual field before a selection is made \citep{Kumar2009}. Therefore, our attention model is very plausible physically. The word attention is mostly associated with a conscious action, but it is also known that attention can be implicit. For example, studies on the effect of emotions on  explicit task-related attention processes are well documented, and showed that implicit attention to marginal information (stimulus)  is possible \citep{Flaisch2015}. In our model, we assume both types of attention mechanisms are involved in a decision task. 

  For this particular example, there are four value states, and all have to be maintained in the working memory simultaneously. This is already a very taxing process because of the limited capacity of the short term memory (STM). It has been suggested that on average, a person can hold concurrently around four to seven chunks of information in
STM \citep{Miller1956,Cowan2001,Aben2012}. The parameter $\omega$  represents the switching rate between the states $ \$14 \Leftrightarrow \$ 90$, while $\epsilon$ is a constant distribution that may represent the degree of bias or belief of the DM towards the decision states. Hence $\omega$ and $\epsilon$ are associated with the WM (or demon). In suggesting this particular circuit, we are assuming that the DM really focuses on states and not gambles. If the most probable state happens to be in gamble A, then gamble A is chosen. Therefore, the mental representation we have adopted is the following \citep{Rescorla2015}. 
The Maxwell demon lives in the DM's brain \citep{Maxwell1891}, and it is the WM of the DM that 'drives' the excited neurons to a certain goal state based on the information just acquired and the subjective information that is already stored in the mind. The demon's job is to settle on a single state as soon as possible given the subjective state of the mind. This process is equivalent to 'erasing' information from the region of the brain that is directly involved in processing the external information. Hence, upon decision, entropy decrease may happen in the WM of the demon, but overall, entropy increases in the whole brain network.


The rate equations associated with this network are
\begin{eqnarray}
\dot{p}_1 & = &  \omega \,\epsilon p_2 - \omega (1 - \epsilon) p_1 + W_{31} p_3 - W_{13} p_1 + W_{41} p_4 - W_{14} p_1,  \nonumber \\
\dot{p}_2  & = & - \omega \, \epsilon p_2 + \omega (1-\epsilon) p_1 +  W_{42} p_4 - W_{24} p_2 + W_{32} p_3 - W_{23} p_2,  \label{masterIJ} \\
\dot{p}_3  & = & W_{23} p_2 - W_{32} p_3 + W_{13} p_1 - W_{31} p_3 + W_{43} p_4 - W_{34} p_3, \nonumber \\
\dot{p}_4  & = & W_{14} p_1 - W_{41} p_4 + W_{34} p_3 - W_{43} p_4 + W_{24} p_2 - W_{42} p_4, \nonumber
\end{eqnarray} 
where the probabilities satisfy $p_1 + p_2 + p_3 + p_4 = 1$, and are not equal to the initial probabilities. The initial (i.e., stimulus) probabilities are now encoded in the transition rates $W_{ij}$ according to Eq. \ref{wij}.

\begin{figure}[H]
	\includegraphics[width=6in,height=4in]{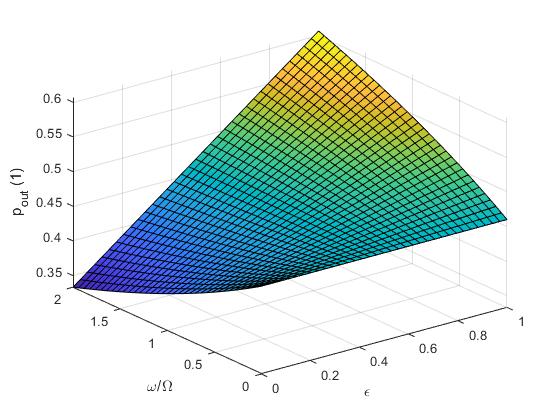}
	\caption{Violation of Stochastic Dominance: The probability of state $\$90$ which corresponds to bit 1 in the memory as calculated from the viewpoint of the memory (demon). The data is for $k_I = k_J = \Omega$. ---}
	\label{IJplot}
\end{figure}

For the working memory,  we adopt the simplest model. The bits 1 and 0 are associated with states $\$90$ and $\$40$, respectively.  The choice of these two states appear arbitrary, but what is important is that they do not belong to the same lottery. Otherwise, their choice does not have a significant effect on the final distribution. 
For simplicity, the incoming bits are assumed to have probability $p(1) = \epsilon$.  The bits interact for a period $T$ with the network according to the rate equations above, Eq. \ref{masterIJ}, before the next bit arrives. 
After the interaction period, the network evolves to a new state such that the value outcome $\$90$ has a new probability, $p_1$. The change in $p_1$ will be reflected in the outgoing part of the tape such that $p_{out}(1) = p_1 /(p_1+ p_2)$.  This system reaches a non-equilibrium steady state with
\begin{eqnarray}
p_{out} ( 1) & = &  \frac{0.46 k_1  + 0.58 k_2 + \omega \epsilon \left(0.21 + 0.12 k_2/k_1 + 0.09 k_1/k_2\right)}{k_1+  1.24 k_2 + \omega \left(0.21 + 0.12 k_2/k_1 + 0.09 k_1/k_2\right) }, 
\label{p1IJ}
\end{eqnarray}
where $k_1$ and $k_2$ are the corresponding transition probability rates for $J$ and $I$, respectively. From reaction rate theory, the ratios of the transition probabilities are expected to be such that $k_1/k_2 \propto \left|\beta_2/\beta_1 \right|^\nu$, with $\nu > 0$ \citep{Hanggi1990}. This gives us an idea of the magnitude of possible ratios which clearly can affect the probabilities of choices made in an important way. If we take this ratio to be 1, the probability of state $\$ 90$ in the memory becomes 
\begin{eqnarray}
p_{out}(1) & = & 0.46 \frac{1+0.41 \epsilon \omega/\Omega}{1+ 0.19 \omega/\Omega},
\label{IJ}
\end{eqnarray}
 where $k_1 = k_2 = \Omega$. The function in Eq. \ref{IJ} is displayed in  Fig. \ref{IJplot}. Therefore, for small ratios of $\omega/\Omega$, state $J$ can never be chosen according to our scheme. In the opposite limit of this ratio, the state $\$ 90$ becomes abundant in the memory, or in other words, attention is more focused on this state than other states. Physically, this limit corresponds to very fast decisions. However, this latter approximation is very poor given that according to the entropic utility solution, the ratio of $\beta_J/\beta_I \approx 50$ which is far larger than 1.  
 Note that according to the limited capacity of the working memory, the DM cannot keep track in the STM of all the options available. Therefore tracking few options in the WM will be the most plausible way that occurs during a decision process.

 While we considered this network solution as occurring within a single brain, this need not be the case. We may equally use each bit to be associated with a different person. Hence in this way the steady solution corresponds to the choice of the overall population, and $p_{out}(1)$ corresponds to the percentage of population choosing gamble J.  In this case, if the population starts from the $\$ 90$ state upon being asked to choose between $I$ and $J$, only about $67 \%$ of them will end up choosing $J$. According to Birnbaum's measurements, $65.8\%$ of the population choose $J$ over $I$ in 12 studies  \citep{Birnbaum2004b}.  

\subsubsection{Allais paradox}

\begin{figure}[H]
	\centering
	\subfloat[Allais AB]{
		\includegraphics[width=0.6\textwidth]{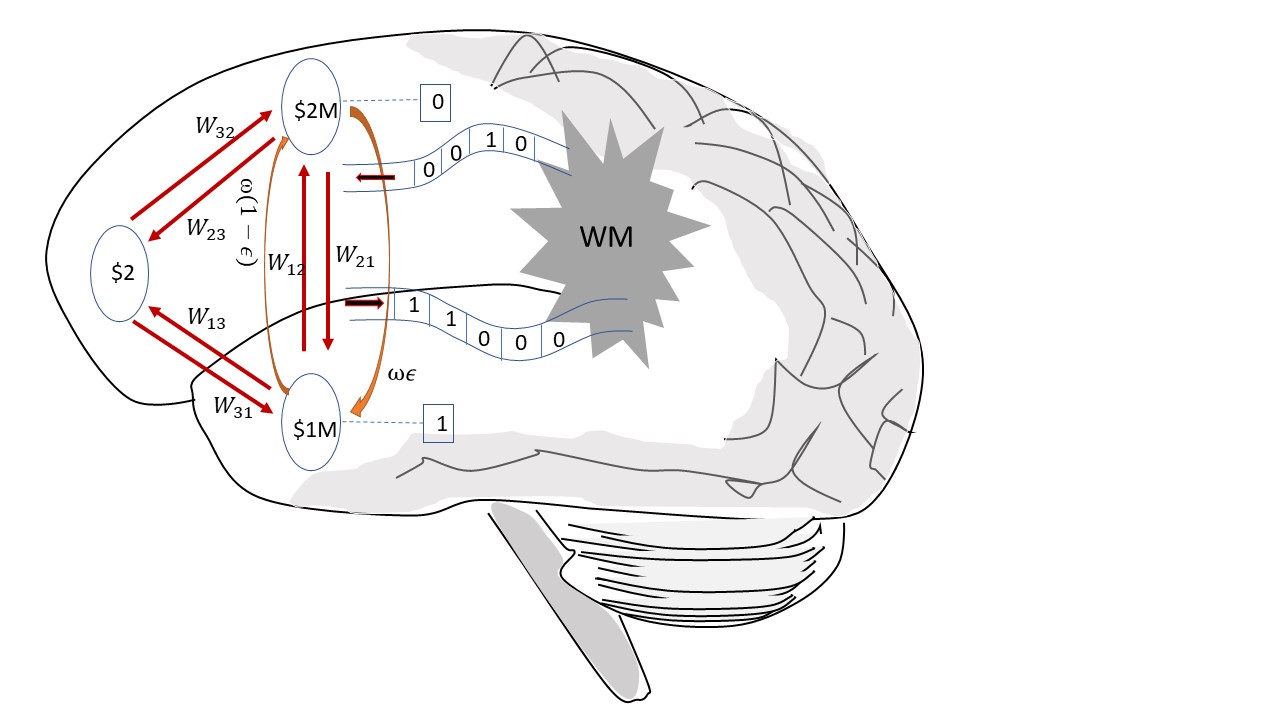}
	}
	\subfloat[Allais CD]{
		\includegraphics[width=0.6\textwidth]{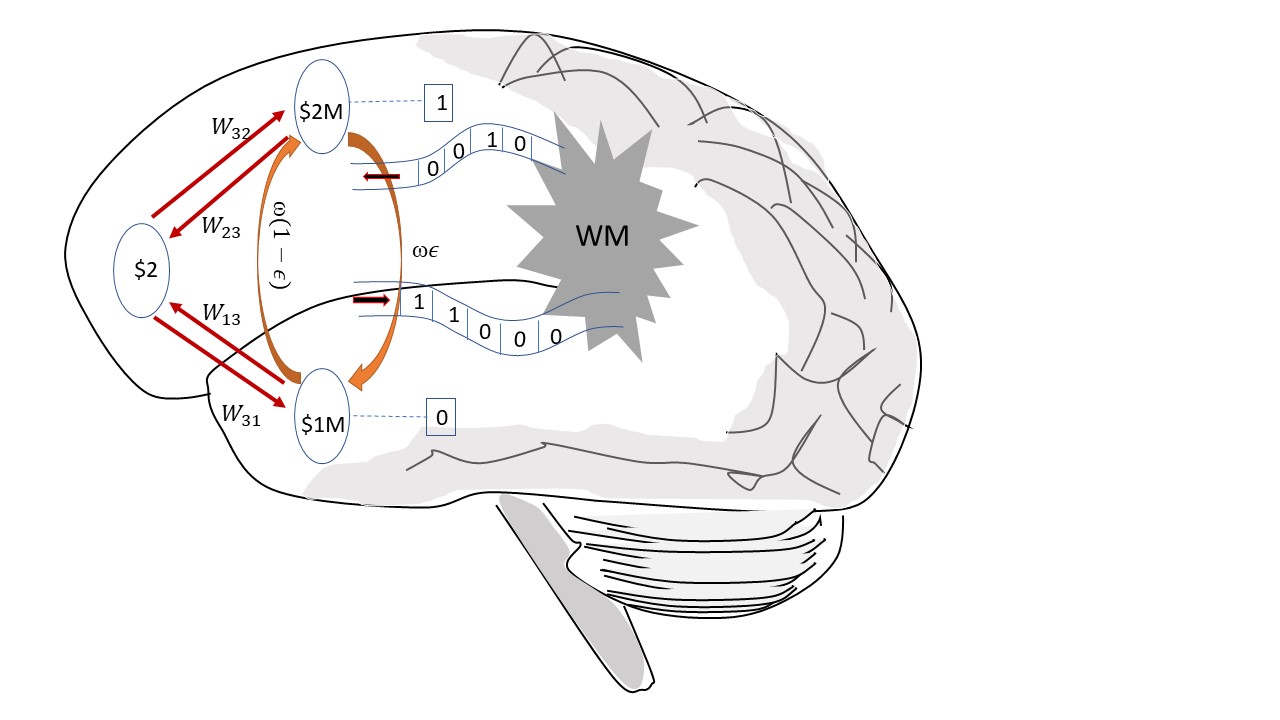}
	}
	\vspace{0.5cm}
	\caption{Allais paradox \citep{Birnbaum2008}: (a) Allais Paradox: A vs. B. The A lottery consists of $\$ 1 M$ option with certainty, while the B lottery has three options with $ (\$ 2, 0.01), (\$ 1 M, 0.89)$ and $(\$ 2 M, 0.10)$.  (b) Allais Paradox: C vs. D. The C lottery consists of $(\$ 1 M, \, 11\%)$ and $( \$ 2, \, 89 \%) $ options, while the D lottery has two options with $ (\$ 2 M, \, 10 \%)$ and $ (\$ 2, \, 90 \%)$. }
	\label{allaisg}
\end{figure}

First,  we treat the pair of gambles $A$ and $B$. The circuit for this case is shown in Fig. \ref{allaisg}-a. Here the transition rate, $\omega$, between the $\$ 2M$ state and $\$ 1M$ state is influenced by the presence of the certain state of $\$ 1M$. Hence, $\omega$ will be proportional to $1/|\beta_A|$. Since people will focus their attention on the relative chances of receiving $\$2M$, a naive estimation is to evaluate $p(\$ 2M)/(p(\$1M) + p(\$2M))$ which is approximately  equal to $10 \%$, the same as in the original $B$ gamble since the $\$2$ state is barely relevant.  Therefore, we expect including the certain state to make receiving $\$2M$  less probable than receiving $\$ 1M$ even in the most pessimistic case that of $\epsilon = 0$. To calculate the chances of choosing the state of $\$ 2M$, and the  gamble $B$, we first write the rate equations for the circuit.

Let $p_1, p_2$ and $p_3$ the probabilities of the $\$ 2M$, $\$ 1 M$, and $\$ 2$ states, respectively. The rate of change of these probabilities for the lotteries A-B are 
\begin{eqnarray}
\dot{p}_1 & = & W_{21} p_2 + W_{31} p_3  - W_{13} p_1  - W_{12} p_1 + \omega \,\epsilon p_2 + \omega (1 - \epsilon) p_1 \nonumber \\
\dot{p}_2  & = & W_{32} p_3 - W_{23} p_2 + W_{12} p_1 - W_{21} p_2 - \omega \, \epsilon p_2 + \omega (1-\epsilon) p_1 \\
\dot{p}_3  & = & W_{31} p_1 - W_{31} p_3 + W_{23} p_2 - W_{32} p_3, \nonumber
\end{eqnarray} 
where $\omega$ is the frequency of transitions between the two lotteries. Therefore it is assumed to reflect the process of attending to the various outcome values.    The outgoing probability of having a bit of 1 in the memory after the coupling between the memory and network that encodes the gambles is
\begin{eqnarray}
p_{out} (1) & = & \frac{p_1}{p_1 + p_2} \nonumber \\
& = & \frac{W_{23} W_{31} + (W_{31} + W_{32}) \left(W_{21} + \omega \, \epsilon\right)}{W_{21} W_{31} +W_{23}W_{31}+ +W_{13}W_{32} + (W_{31} + W_{32}) (W_{12}  + \omega )}. \nonumber \\
&& \nonumber \\
 & = & 0.898 \,\, \frac{1 + 0.31 \, \epsilon \,\, \frac{\omega}{k_B}}{1 + 0.28 \, \,  \frac{\omega}{k_B}}.
\end{eqnarray}
Thus according to this representation, the relative probability of  the $\$1M$ state compared to the $\$ 2M$ state in the registered memory is at worst $75\%$, with complete bias toward the $\$ 2M$ state, i.e., $\epsilon = 0$ and $\frac{\omega}{k_B} \approx \frac{1}{2}(k_A+k_B)$. The treatment discussed here can be interpreted differently. If we assume that each time step corresponds to a different DM, and that the $\$ 1 M$ state attracts the attention of each DM the same way. We estimate in this case that only $8\%$ of the population will choose to play lottery $B$ and not $A$. This estimate is very close to the one reported in \citet{Kahneman1979}, about $12\%$,  but far smaller than the $42\%$ measured by \citet{Birnbaum2008}. This latter experiment was not part of his extensive studies reported in \citet{Birnbaum2004b}.

The circuit for the two lotteries $C$ and $D$ in the brain is schematically  represented in Fig. \ref{allaisg}-b. Similar to previous set of lotteries, we  write the rate equations for the $CD$ network in the value-based space of choices. 
The rate of change of probabilities in this coupled system are governed by
\begin{eqnarray}
\dot{p}_1  & = & W_{31} p_3 -W_{13} p_1 - \omega \, \epsilon p_1 + \omega (1-\epsilon) p_2 \nonumber \\
\dot{p}_2 & = & W_{32} p_2 - W_{23} p_2 + \omega \, \epsilon p_1 - \omega (1-\epsilon) p_2 \\
\dot{p}_3  & = & W_{13} p_1 - W_{31} p_3 + W_{23} p_2 - W_{32} p_3. \nonumber
\end{eqnarray}
The probability of bit 1 in the memory after many interaction periods with the network representing both  gambles is 
\begin{eqnarray}
p_{out} (1) & = & \frac{p_2}{p_1 + p_2} \, =\, \frac{W_{13} W_{32} + \omega \,\epsilon (W_{31} + W_{32} )}{W_{23} W_{31} + W_{13} W_{32} +\omega  (W_{31} + W_{32})}.
\end{eqnarray}
Using the probability transition rates, we have
\begin{eqnarray}
p_{out} (1) & = & \frac{  0.92 +  \,\, \epsilon \, \frac{\omega}{k_C }\left( \,\,\frac{1}{3} + 0.365\,\,  \frac{k_C}{k_D} \,  \right)}{  2.01 +  \, \, \frac{\omega}{k_C} \left(  \,\, \frac{1}{3} + 0.365\,\,  \frac{k_C}{k_D} \,  \right)}.
\label{allaiscd}
\end{eqnarray}
For this pair of lotteries, \citet{Birnbaum2008} gave $76\%$ as the percentage of people who chose lottery $D$ over lottery $C$. This pair of gambles were not part of the 12 studies we quoted above. To get a percentage close to the measured value, the population must have a bias toward the $\$2M$ state, i.e., $\epsilon > 0.5$, and for average interaction periods between the WM and the network much shorter than the transition times between the outcomes of the $D$ gamble. The requirement of bias is not necessarily a weakness of the model, but is considered a prediction in this instance. Diffusion-based models  \citep{Ratcliff2016}, for-example, require a bias parameter that needs to be varied to achieve agreements in two-choice decision tasks. From Table \ref{tableallais}, we see that the focus of attention, i.e., $1/k_D$ is much longer than that of gamble $C$. Therefore, if the interaction period between memory and network is much larger than the average transition times between outcomes of gamble $C$, but less than that of gamble $D$, we can attain probabilities in the range measured by the experiment. These are details that can be checked in future measurements.

\subsubsection{Birnbaum Paradoxes}

Next we suggest an interactive circuit for the Birnbaum paradoxes.

\paragraph{Upper Cumulative Independence}

\begin{figure}[H]
	\includegraphics[width=6in,height=4in]{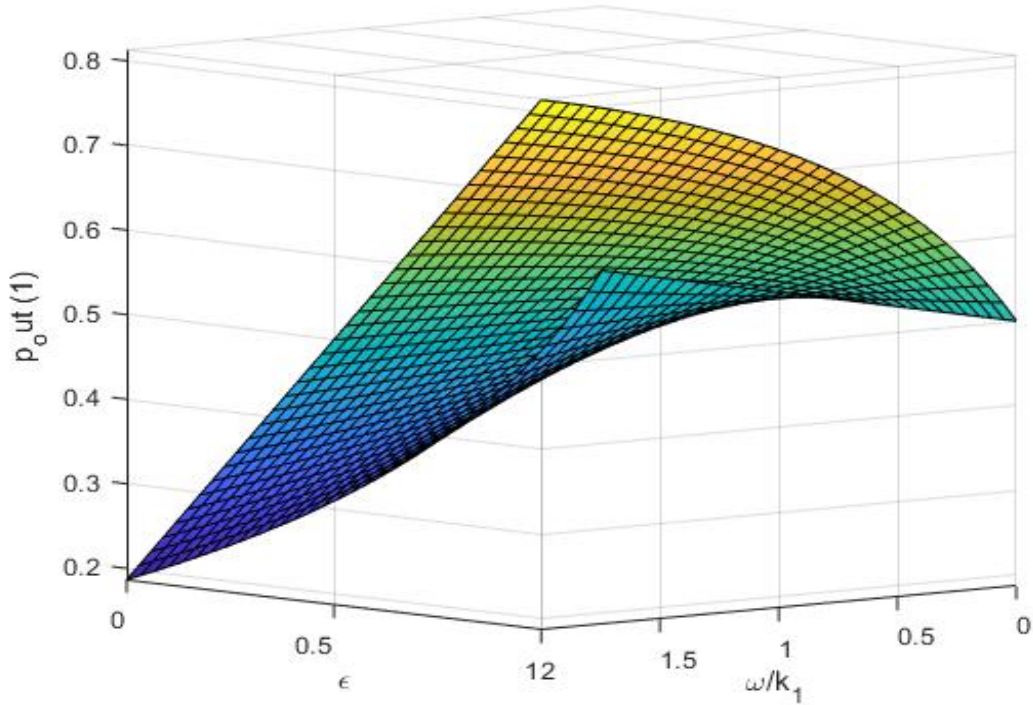}
	\caption{Upper cumulative independence in the interactive model: $R'$ versus $S'$ of test no. 10 in Table 3 of \citet{Birnbaum2004b}. The data is for $k_{S'}/ k_{R'}= 5$ . --- }
	\label{graphuc1}
\end{figure}

For the upper cumulative independence test, we use the pair of gambles 10 and 9 in Table 3 of \citet{Birnbaum2004b}. As we observed above, we will interpret our results from two angles, that of a single DM or a population of DMs. 
\begin{figure}[H]
	\centering
	\subfloat[R'S']{
	\includegraphics[width=0.7\textwidth]{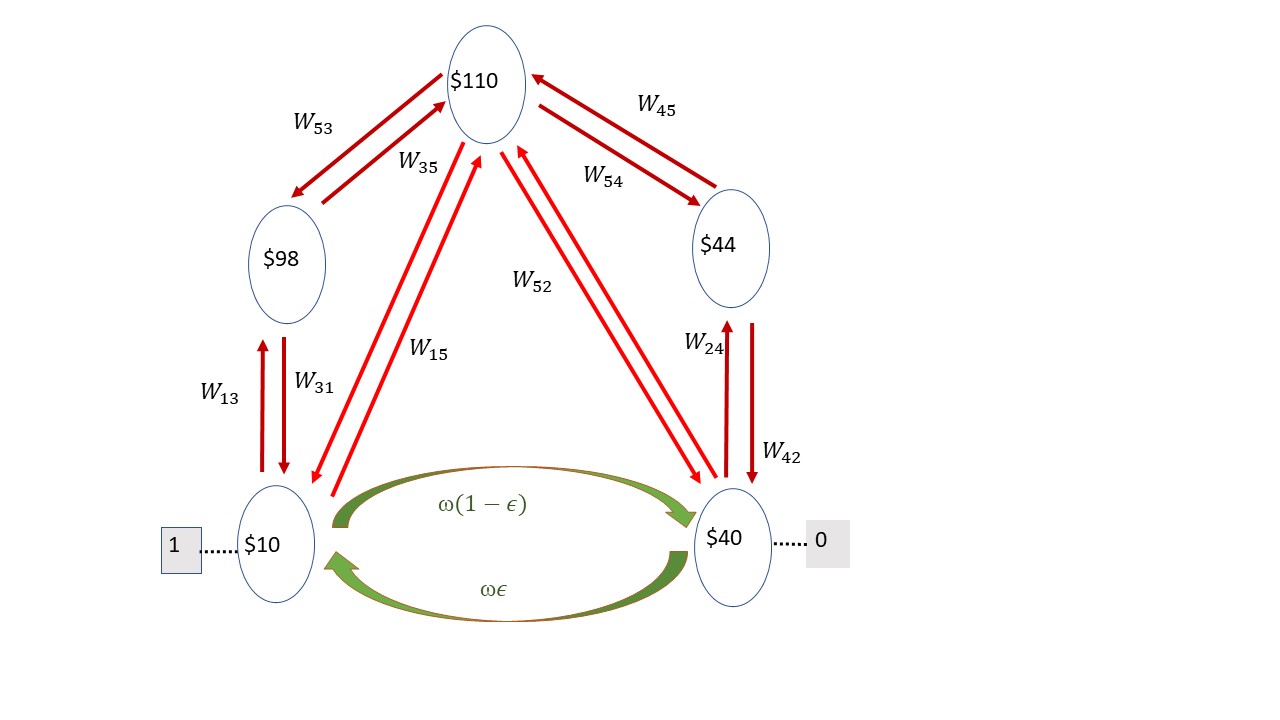}
}
\subfloat[R'''S''']{
	
		\includegraphics[width=0.7\textwidth]{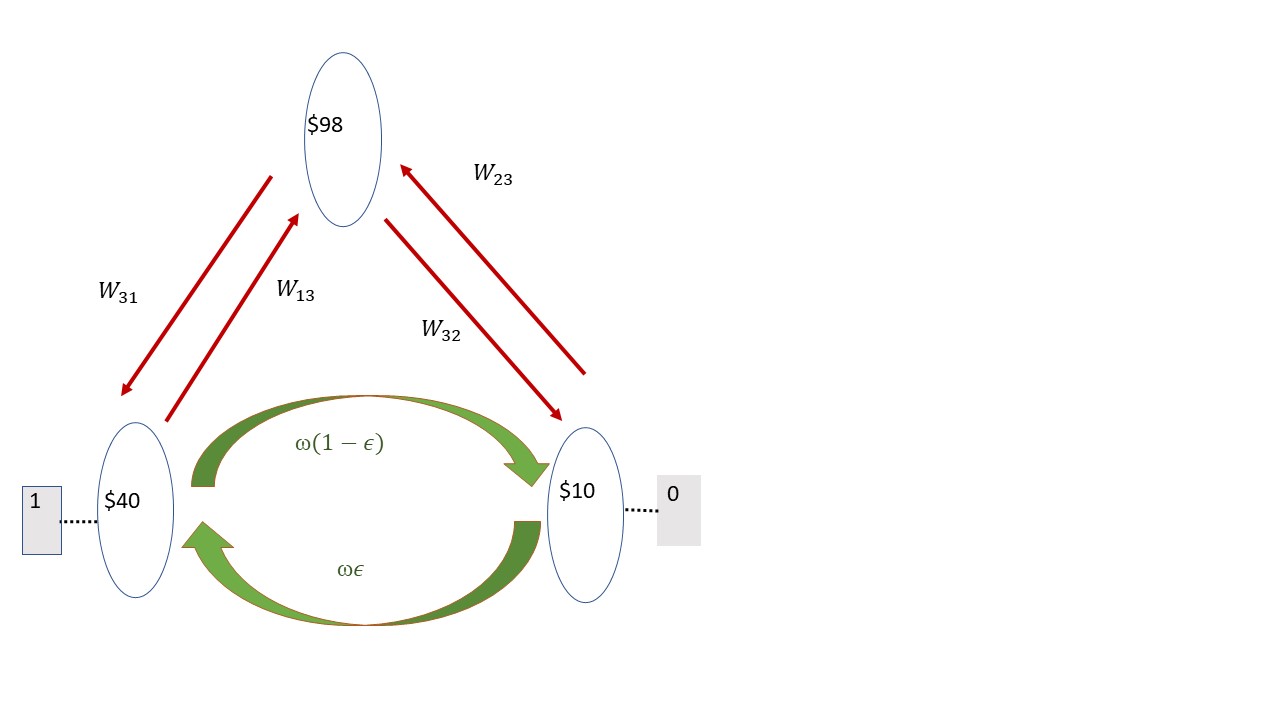}
	}
	\caption{Upper cumulative independence: (a)  $R'$ versus $S'$  choice 10 in Table 3,
		(b) $S'''$ versus $R''''$ choice 9 in Table 3 in \citep{Birnbaum2004b}}
	\label{uc109}
\end{figure}
The initial probability of bit 1 is $p_{in}(1) = \epsilon$. After interaction with the bit, the new probability of bit 1 in the $R'S'$ circuit becomes
\begin{eqnarray}
p_{out}(1) & = & \frac{0.5 + 0.14 \, \epsilon \,\, \frac{\omega}{k_1} \left(1 + \frac{k_1}{k_2} \right)}{1 +  0.14 \, \frac{\omega}{k_1} \left( 1 + {k_1}{k_2} \right)}
\label{pr1}
\end{eqnarray}
This functional dependence is plotted in Fig. \ref{graphuc1} for different values of the initial distribution of the bits 0 and 1, and how fast memory manipulation is being carried out. Clearly, in this case for $\epsilon = 0.5$, the attention scheme does not work since it can't decide on its own which state to choose. A slight bias is needed for a choice to be made.  According to the measurements of \citet{Birnbaum2004b} averaged over his 12 studies, $69\%$ of people chose gamble $R'$ over gamble $S'$. If we adopt the view that every interaction interval between the demon and the gambles are independent and correspond to different people, the probability of outgoing bit 1 should be closely aligned with the percentage of people choosing $R'$ in the limit of an infinite sample of people. Therefore, according to Eq. \ref{pr1}, to achieve comparable percentages, the population should have an initial bias toward the higher value state of $\$ 98$ in lottery $R'$. As an example, if we take $k_1/k_2 \approx \beta_{R'}/\beta_{S'} \approx 5$, Table \ref{tableUCI}, and choose the rate of information manipulation by the WM to be approximately $\omega/k_1 =2/5$, i.e., midway between the intra-transition rates, we find $p_{out}(1) \equiv 0.63$.   

A similar calculation for the second pair of gambles, $R'''S'''$ of test 9 in \citet{Birnbaum2004b} gives for the probability of outgoing bit 1 after interaction with the network of the gamble
\begin{eqnarray}
p_{out} ( 1 ) & = &  \frac{1.5 +  \, \,\epsilon \,\, \omega/3k_2 \left( 1 + 1.5\,  k_2/k_1 \right)}{2.17 +  \, \, \omega/3k_2 \left( 1 + 1.5 \, k_2/k_1  \right)}.
\label{pr2}
\end{eqnarray}

\begin{figure}[H]
	\includegraphics[width=7in,height=6in]{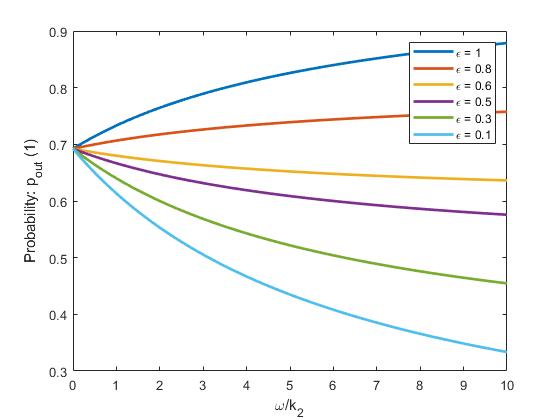}
	\caption{Violation of upper cumulative independence: $S'''\, vs \, R''''$ choice 9 in Table 3 in \citep{Birnbaum2004b}. The data is for $k_{S'''}/k_{R'''} = 0$.}
	\label{graphuc2}
\end{figure}

This function is shown in Fig. \ref{graphuc2} for various values of the initial bit 1. The percentage value of the population that chose $S'''$ over lottery $R'''$ averaged over 12 studies was about $63\%$. Such percentage can be understood either as a bias toward the $\$40$ state or faster shifting of attention between gambles than that within gamble $S'''$. The latter means that more attention is paid to the chosen gamble.

 The plot is shown for the case when $k_{S'''}/k_{R'''} \approx 0$ , which is implied by entropic utility, Table \ref{tableUCI2}. This result is physically plausible. The parameters $\omega$ and $\epsilon$ reflect cognitive effort by the WM, this implies that the memory can interact very little with the network and induce a decision. Therefore, compared to the decision made on the $R'S'$ pair, it is  easier to make a decision on this pair. Again, this is very reasonable given the complexity of the first pair. Since the measured probabilities of the first decision are higher than this current one, this may lead us to believe that it is easier to make a decision on the first pair rather than this one. From the structure of both pairs, it is very difficult to agree with this interpretation, and we believe that the conclusion from this model is more plausible in this case.

\paragraph{Lower Cumulative Independence}

Finally, we address the violation of lower cumulative independence with this transfer of attention approach, and discuss how this violation can be understood from the WM memory manipulation of the regions encoding the stimulus, i.e., the gambles. Lower cumulative independence implies that if people prefer $S$ over $R$, then they should also prefer $S''$ over $R''$.

\begin{figure}[H]
	\centering
	\subfloat[RS]{
	\includegraphics[width=0.5\textwidth]{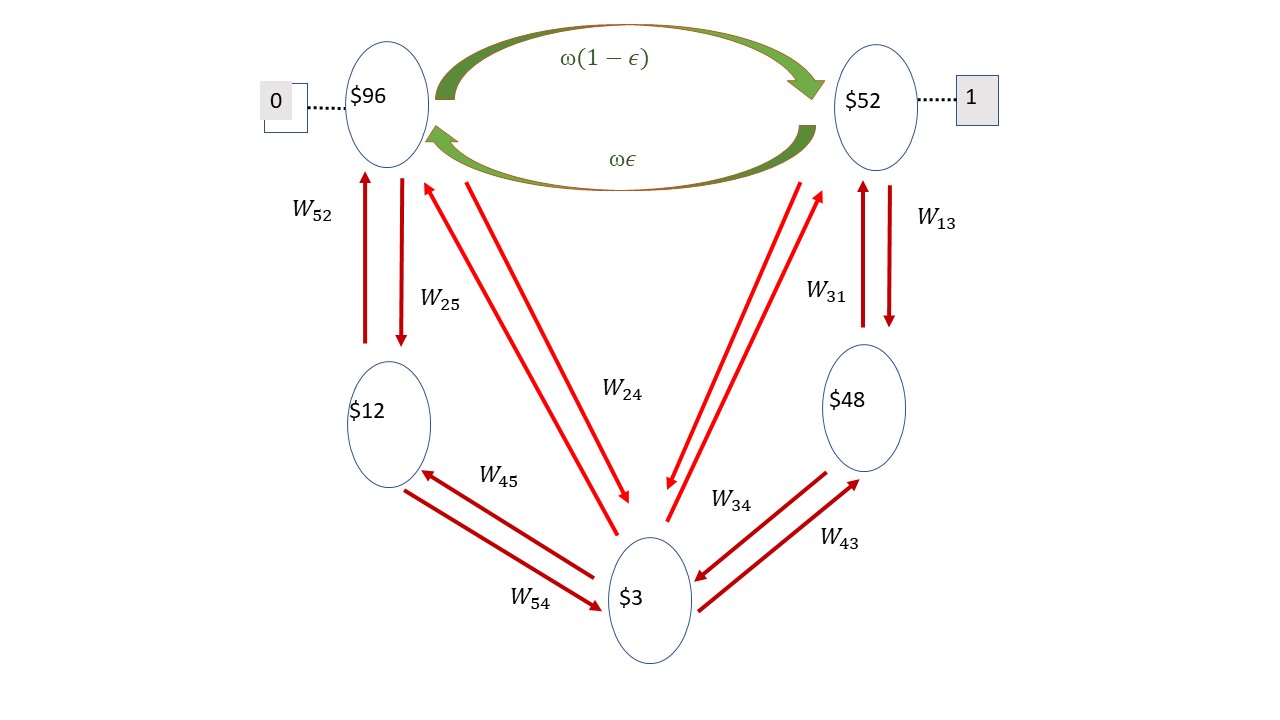}
}
\subfloat[R''S'']{
	\includegraphics[width=0.5\textwidth]{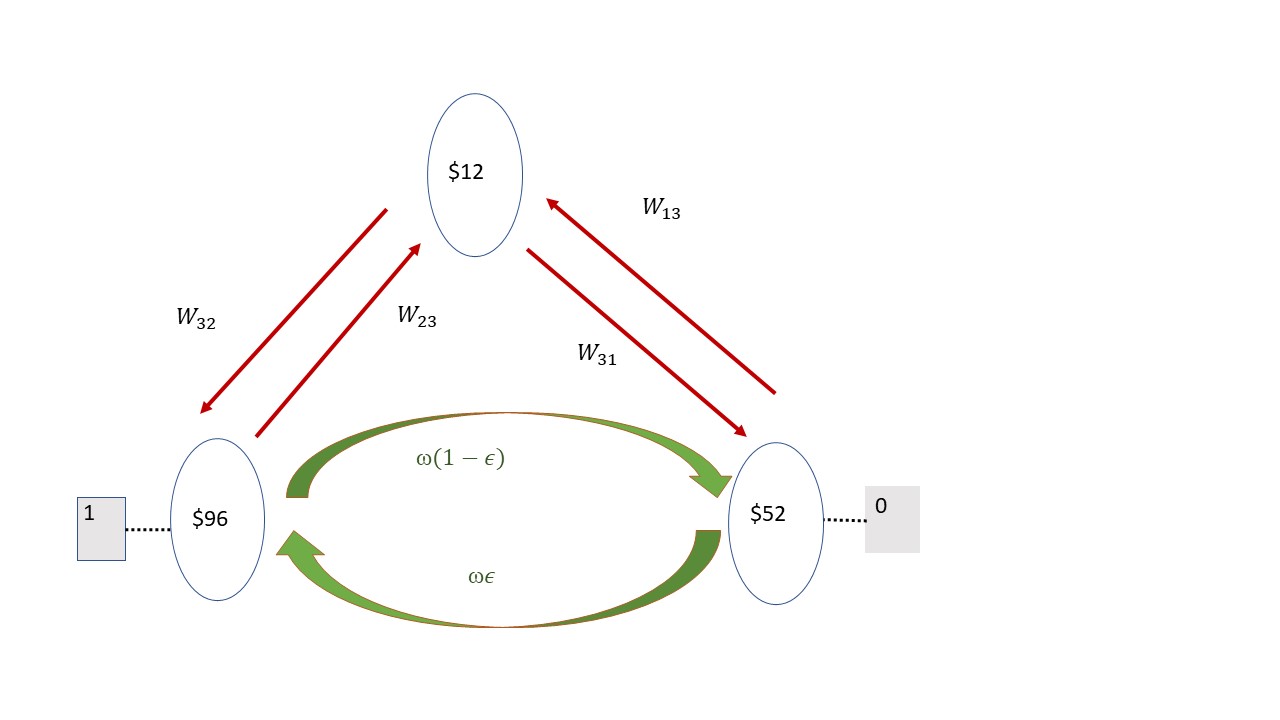}
}
\vspace{0.6cm}
	\caption{Lower cumulative independence circuits: (a)   R versus S lotteries: Table 3 , test no. 17 in \citep{Birnbaum2004b}      (b) R'' versus S'' lotteries: Table 3, test no. 20 in  \citep{Birnbaum2004b}).  }
	\label{RvsS_circuit}
\end{figure}

For lower cumulative independence, Birnbaum  studied the following pair of lotteries in studies 17 and 20 of his Table 3 \citep{Birnbaum2004b}. The gambles are $R = (\$ 3, 0.90; \, \$12, 0.05; \, \$96, 0.05)$ and $S = ( \$3, 0.90; \, \$48, 0.05; \, \$52, 0.05)$ for the first pair of choices, and $R'' = (\$12, 0.95; \$96, 0.05)$ and $S'' = ( \$12, 0.90; \, \$ 52, 0.10)$ for the second pair of gambles. The circuits for the two pairs of gambles are shown in Fig. \ref{RvsS_circuit}. 

For the pair of gambles $RS$, the probability of having bit 1 in the memory after interaction with the network is given by 
\begin{eqnarray}
p(S) & = & 0.5 \frac{1 + 0.20 \,\epsilon\, \frac{\omega}{k_1} (1 + \frac{k_1}{k_2})}{1 + 0.10 \frac{\omega}{k_1} ( 1 + \frac{k_1}{k_2} )}. 
\end{eqnarray}
The parameters of the entropic utility for this pair of gambles suggests that the ratio $k_1/k_2 \approx 3$ 

\begin{figure}
	\includegraphics[width=\textwidth]{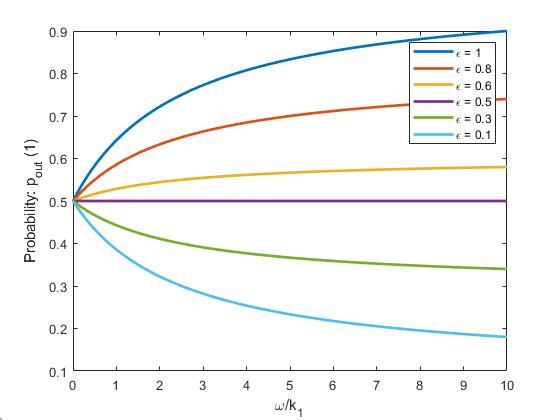}
	\caption{Lower cumulative independence: R vs S lotteries. Probability of choosing R as a function of the parameters $\omega$, $k_1$, and $\epsilon$ (\citep{Birnbaum2004b}, test 17). The data is for $k_1/k_2 = 3$. ------}
	\label{lci17}
\end{figure}

According to Birnbaum 's measurements, there was approximately $52\%$ of the population who chose $S$ over $R$. According to the suggested circuit, a preference of $S$ over $R$ will need at least some bias towards the state $S$, which is shown in Fig. \ref{lci17}. Moreover, this bias will be accentuated for higher $\omega/k_S$ ($k_1=k_S$), i.e., when the WM interacts with the S-side of network more often than with R. This is again a plausible result which means that most of the attention is focused on the $S$ gamble, and the more attention a gamble is attended to, the more often  it is chosen. Hence, in spite of the simplicity of the model, the details of the  prediction is something that can be tested.

\begin{figure}
	\includegraphics[width=\textwidth]{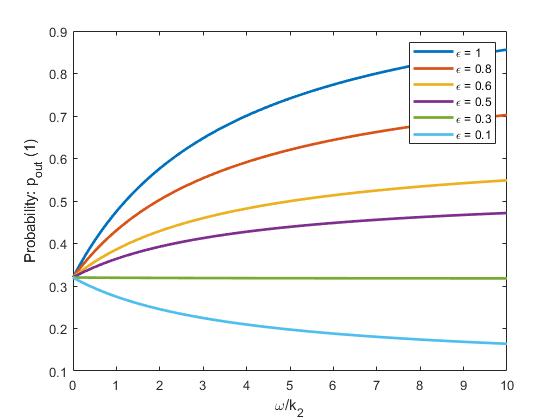}
	\caption{Lower cumulative independence: $R''$ vs $S''$ lotteries. Probability of choosing R as a function of the parameters $\omega$, $k_1$, and $\epsilon$ (\citep{Birnbaum2004b}, test 20). The data is for $k1/k_2 = 1$. --}
	\label{lci20}
\end{figure}

For the second pair of gambles $R''$ and $S''$ in test 20 of \citet{Birnbaum2004b}, the probability of the bit 1 in the memory after the interaction with the network is 
\begin{eqnarray}
p(R'') & = & 0.321 \,\, \frac{1 +  \, \epsilon \,\, \frac{\omega}{k_2}(0.485 + \frac{k_2}{3k_1})}{1 + \,\,  \, \frac{\omega}{k_2}(0.156 + 0.11 \frac{k_2}{k_1})}.
\end{eqnarray} 
Here, the choice of  $R''$ depends heavily on the ratio $\omega/k_2$ in addition to an initial bias towards the largest value state of $\$ 96$. From the energetic picture we tried to lay out in previous sections, it is very plausible that this state will have more motivational momentum because of its high variance, but since this simple attention model cannot quantify the dependence of the rates on the parameter $\lambda$, we cannot study the behavior due to this terms in detail. The dependence of the rates on the variance term has not been included in the previous applications, and therefore our discussion remains incomplete. But in spite of this shortcoming, the  method is still able to shed some light on the interaction of the WM with the various prospects of the lotteries. The data is shown in Fig. \ref{lci20}. 

Birnbaum's measurements for this pair of gambles was about $69\%$. For this particular pair of gambles, the attention is mainly concentrated on gamble $R''$ and little attention is paid to gamble $S"$. In addition to the bias, the reaction time of the WM with the lotteries must be less than that of gamble $R''$. Therefore, attention is more concentrated on the chosen item similar to earlier conclusions. 

The advantage of including a component representing WM as part of the decision problem is obvious from these simple illustrations. Even though our discussion involved only the qualitative influence of the $\beta$ parameter which is related to the average anticipated utility, we completely ignored the direct effect of the parameter $\lambda$ on the attention process. As we demonstrated in the previous section, this term is essential to the decision problem and cannot be ignored. However, its effect will be difficult to quantify theoretically.


\vspace{2cm}

\section{CONCLUSION}

Much progress has been made in identifying regions of the brain that are involved in decision making under risk. However, little is known about the actual mechanisms of choice. Normative theories of decision making have had very limited success in accounting for many anomalies that appear to be irrational from the viewpoint of the laws of probability. But, descriptive theories such as Prospect Theory were able to address some of these anomalies by introducing a non-linear weighting function of the probabilities in the utility function. Unfortunately,  Prospect Theory turned out to be itself incomplete. For example, the TAX model of Birnbaum is more successful than  PT in dealing with violations of coalescence.  Though ultimately, a good theory of decision making will be a theory that is founded on physical principles. The literature we tried to review in this paper already point to the need to treat information and energy in a unified manner to have a consistent picture of entropy that is valid both at the behavioral and neuronal dynamic levels. In this work, by assuming that values in a gamble are proportional to some energy scale, we were able to show that energy and information can be combined in a consistent manner where entropy is maximized but subject to some constraints. We provided three different angles from which a decision making problem can be approached using the same principle of entropy maximization. 

In the first approach, we showed how a coarse grained picture of spikes, that encode probabilities of value (energy) states and its variance, can describe fluctuations that affect the average value of the gamble and hence influence the decision process. The decision in this case is driven by the gamble that has the best signal to noise ratio. Battleground noise in the OFC is significant and affect spike rate of neurons in this region which is the region where encoding is believed to be carried out.

In the second approach, a more macroscopic picture is adopted. But, instead of fluctuations in probabilities, a non-linear description of the decision process is introduced that includes taking account of the  variance of the value states. This allowed us to introduce the entropic utility function, which we argued that it can be understood as a sum of anticipated gain and   information cost associated with the gain function. Maximization of entropy is introduced here through Jaynes' principle of non-extensive entropy, which is necessary to account for various cases.

Finally, the third approach is an attempt to use rate equations to simulate attention as a way to process the different options of both gambles simultaneously and to arrive at a decision. This approach is dynamical but we treated only the steady state solution which 
can be interpreted as the choice made by an infinitely large population. We also showed the importance of including an additional subsystem that of working memory to represent choice as  a competition that may be affected by the state of mind of the DM. This approach has its own merits since it is capable of discussing entropy production,  a topic that we do not discuss but might be important in the details of a dynamical theory of decision making. Therefore, this approach provides a complementary physical picture to the entropic discussion in the second approach.

In all three approaches, maximization of entropy is respected. 
While we expect all three approaches to be complementary and part of a complete decision theory, there is still the important unanswered question of how a decision is actually carried out. We know from microscopic dynamics that inhibiting neurons play an important role in the dynamics of spiking neurons. Therefore, including the physics of inhibiting neurons is essential for a more physical model of decision making. Throughout this paper, we tried to treat the dynamics of decision making only in a qualitative manner and point to the importance of a non-equilibrium (critical) state for decision making to happen. Decision making, in this picture, is very similar to a self-organized critical behavior which is another reason why non-extensive entropy is the proper entropy to use in this problem since it allows us to get observable scaling laws \citep{Kello2010}.  \\



\end{document}